\documentclass[aps,floatfix,prd,superscriptaddress,twocolumn]{revtex4-1}

\usepackage{amsmath}              
\usepackage{amssymb}              
\usepackage{braket}               
\usepackage[mathscr]{euscript}    
\usepackage{graphicx}             
\usepackage{hyperref}             
\usepackage{listings}             
\usepackage{tabularx}             

\begin{document}

\widetext
\leftline{Compiled \today}
\leftline{To be submitted to PRD}

\title{Elastic Scattering in General Relativistic Ray Tracing for Neutrinos}

\author{M.\ Brett Deaton}
\email{mbdeaton@ncsu.edu}
\affiliation{Joint Institute for Nuclear Astrophysics,
  Michigan State University, East Lansing, MI 48824, USA}
\affiliation{Department of Physics,
  North Carolina State University, Raleigh, NC 27695, USA}

\author{Evan O'Connor}
\affiliation{Department of Astronomy and Oskar Klein Centre,
  Stockholm University, Alba Nova, SE-10691, Sweden}
\affiliation{Department of Physics,
  North Carolina State University, Raleigh, NC 27695, USA}

\author{Y.\ L.\ Zhu}
\affiliation{Department of Physics,
  North Carolina State University, Raleigh, NC 27695, USA}

\author{Andy Bohn}
\affiliation{Center for Radiophysics and Space Research,
  Cornell University, Ithaca, NY 14853, USA}

\author{Jerred Jesse}
\affiliation{Department of Physics \& Astronomy,
  Washington State University, Pullman, WA 99164, USA}

\author{Francois Foucart}
\affiliation{Department of Physics,
  University of New Hampshire, Durham, New Hampshire 03824, USA}
\affiliation{Lawrence Berkeley National Laboratory,
  1 Cyclotron Rd., Berkeley, CA 94720, USA}

\author{Matthew D.\ Duez}
\affiliation{Department of Physics \& Astronomy,
  Washington State University, Pullman, WA 99164, USA}

\author{G.\ C.\ McLaughlin}
\affiliation{Department of Physics,
  North Carolina State University, Raleigh, NC 27695, USA}


\begin{abstract}
  We present a covariant ray tracing algorithm for computing high-resolution
  neutrino distributions in general relativistic numerical spacetimes with
  hydrodynamical sources.
  Our formulation treats the very important effect of
  elastic scattering of neutrinos off of nuclei and nucleons
  (changing the neutrino's direction but not energy)
  by incorporating estimates of the background neutrino fields.
  Background fields provide information about the spectra and intensities of
  the neutrinos scattered into each ray.
  These background fields may be taken from a low-order moment simulation
  or be ignored, in which case the method
  reduces to a standard state-of-the-art ray tracing formulation.
  The method handles radiation in regimes spanning optically thick to optically thin.
  We test the new code, highlight its strengths and weaknesses,
  and apply it to a simulation of a neutron star merger
  to compute neutrino fluxes and spectra,
  and to demonstrate a neutrino flavor oscillation calculation.
  In that environment, we find qualitatively different fluxes, spectra,
  and oscillation behaviors when elastic scattering is included.
\end{abstract}

\maketitle

\section{Introduction}
Neutrinos are one of the dominant energy transport phenomena at play in
neutron star mergers: heating, cooling, and pushing the disrupted nuclear
matter.
In addition, they change the composition of the matter via charged current
interactions.
Because neutrinos scatter over length scales both large and small with
respect to fluid scales, accurate models require a neutrino treatment that
respects the freedom of neutrino distribution functions to vary
drastically from geometrically simple distributions in thermodynamic equilibrium.

This is a challenging task in the environment of a merger,
which generally lacks any spatial symmetries, so that fully general
solutions to the Boltzmann Equation are not feasible.
Leakage approximations
\citep{ruff1996-leakage_part1,ross2003-leakage_part2,
  deat2013-leakage, gale2013-gr_rad_trans, pere2016-asl, radi2016-dynamical}
capture some of the qualitative effects of neutrinos on the matter, but
provide extremely limited information about the neutrino field itself.
Monte Carlo methods like those
used in supernova simulations \cite{abdi2012-monte_carlo}
and stationary models of accretion disks \cite{rich2015-monte_carlo}
are an excellent tool but require large computational resources
because of the need to use a large number of particles to fully and precisely
sample the high dimensional parameter space
(seven-dimensional in the general case).

The state of the art today for neutron star merger calculations
couples radiation and matter using a truncated moment formalism
\citep{fouc2015-m1_nsbh, fouc2016-m1_nsns},
evolving the zeroth- and first-angular moments of the energy density,
that is the total energy- and momentum-densities.
This method, commonly called M1 transport, was formulated by
\cite{thor1981-truncated_moment} and modernized by
\cite{shib2011-truncated_moment}.
M1 transport has recently become popular in the core-collapse supernova
community as well
\cite{kuro2012-m1_ccsne,just2015-m1_code,ocon2015-gr1d_with_nu,
  skin2016-ray_vs_m1,robe2016-gr_m1_ccsne}.

For merger simulations, \cite{fouc2016-m1_evolve_n} have recently
expanded their M1 transport code
to also evolve the zeroth-angular moment of the number density,
providing total energy densities and average energies throughout the
simulated volume.
Even so, with M1 transport codes only evolving two angular moments
and one or two energy moments,
they can extract only limited angular and spectral information
from the neutrino fields.

But many interesting unsolved problems require an accurate model of the neutrino
spectra and angular distributions.
With a model of the neutrino emission from a merger we can
1) examine neutrino effects on the nucleosynthesis of the ejected material
\citep{surm2011-nickel_56, robe2016-sph_nu_nucleo},
2) explore the rich flavor oscillation physics likely to occur
\citep{malk2012-mnr_1, malk2015-mnr_2, malk2016-mnr_3, zhu2016-mnr_nsns_remnant,
  vaan2016-uncovering_mnr},
3) improve closure relations used in M1 transport schemes
\citep{ramp2002-truncated_moment, shib2011-truncated_moment,
  card2013-truncated_moment, fouc2015-m1_nsbh, ocon2015-gr1d_with_nu}, and
4) study possible jet formation due to neutrino annihilation
\citep{ruff1999-nunubar_nsns, asan2000-nunubar, birk2007-nunubar,
  hari2010-gr_nunubar_collapsar, zala2011-nunubar, leng2014-nunubar,
  just2016-nsns_ejecta_obstacles}.

Angular and spectral neutrino distributions in neutron star merger simulations
have just recently become available with a coupled Monte-Carlo-M1 scheme
\citep{fouc2017-m1_with_mc,fouc2018-mc_for_errors}.
In this work, however, we present a ray tracing method to compute neutrino
distribution functions from the more widely available state-of-the-art
general relativistic M1 transport hydrodynamics simulations.
We choose ray tracing because it is conceptually simple,
numerically inexpensive,
and extends to high resolution in energy and angle
by simply increasing the number of rays sampled.
Furthermore, the computational implementation parallellizes trivially.

With a ray tracing method we approach radiation transport from the perspective
of a single observer at a spacetime event $x_o^\alpha$.
Our goal is to compute the distribution function
$f^{\nu_\sigma}(x_o^\alpha;p_\beta)$, or the amount of neutrino radiation
of species
$\nu_\sigma\in\{\nu_e,\nu_\mu,\nu_\tau,\bar{\nu}_e,\bar{\nu}_\mu,\bar{\nu}_\tau\}$
with momentum $p_\beta$ impinging on $x_o^\alpha$.
To do so we trace a geodesic trajectory from $x_o^\alpha$ in the backwards
direction $-p_\beta$ to sample the incoming radiation along that line of sight.
By tracing a family of rays intersecting $x_o^\alpha$ we build up a
picture of the distribution function there.
And by sampling many observation points we construct a global picture
of $f^{\nu_\sigma}(x^\alpha;p_\beta)$.

The ray tracing framework
is conceptually simple because it solves the equation of radiation
transport (Eqn.~\ref{eqn:boltzmann}) along characteristics,
reducing it to a one-dimensional ordinary differential equation.
It is numerically cheap because it confines computations of
$f^{\nu_\sigma}(x_o^\alpha;p_\beta)$ to the past light-cone of $x_o^\alpha$,
with the history of that light-cone truncated at large optical depth.
It easily extends to high resolution in energy and angle by simply
increasing the number of rays sampled.
And it parallelizes trivially by ray, since each ray is computed
independently.

Several ray tracing formulations for radiation transport already exist.
Most formulations assume an analytic spacetime metric
\citep{birk2007-nunubar, caba2009-detecting_grb_nu,
  hari2010-gr_nunubar_collapsar, kova2011-gr_ray_tracing}.
And many make the simplifying assumption of blackbody emission from a
neutrinosurface
\citep{birk2007-nunubar, caba2009-detecting_grb_nu, kova2011-gr_ray_tracing},
limiting them to equilibrium, optically-thick configurations.
Current state-of-the-art ray tracing formulations avoid the assumption of
blackbody spectra by integrating a local emissivity along each geodesic
(e.g. \cite{hari2010-gr_nunubar_collapsar} for neutrinos and
\cite{youn2012-gr_radiative_transfer} for photons).
But no formulations to date account for the important scattering and pair
processes outlined in Tab.~\ref{tab:neutrino_processes}.
We build upon these existing ray tracing formulations
by eschewing any assumptions about the spacetime geometry,
integrating local emissivities,
and including elastic scattering in the integration along each geodesic.

We formulate the ray tracing equations covariantly---free from
assumptions about the spacetime geometry or coordinates. This is essential
because we want to apply the method as a postprocessing step using
time snapshots of data computed from time-dependent general relativistic evolutions.
The spacetime represented in these snapshots is not analytic (i.e. Kerr).
And even in configurations that are described well by the Kerr metric
(e.g. a low-mass disk around a massive black hole),
the evolution coordinates are unlikely to present the metric in
a familiar analytic form.
This is because integrating the Einstein Equations often requires complicated,
time-dependent gauge conditions
\citep{lind2007-gen_harmonic, fouc2013-compactness_and_spin}.

Elastic scattering (see Tab.~\ref{tab:neutrino_processes}) can signicantly
modify neutrino distributions in angle and dilute the emitted spectrum over a
larger emitting surface \citep{pere2016-asl}.
This is especially pronounced in the case of heavy-lepton neutrinos.
Inelastic scattering and pair processes can introduce further modifications.
Any phenomena that involve neutrino-neutrino interactions,
for example neutrino oscillations
\cite{duan2010-collective,malk2012-mnr_1,vlas2018-multiangle}
and neutrino-antineutrino annihilation \cite{asan2000-nunubar},
depend sensitively on the angular distribution.
And spectral changes in neutrino distributions can strongly affect the nuclear
processes occuring in the ejected and irradiated material
\cite{surm2006-grb_nucsynth,malk2012-mnr_1,caba2012-nu_spectra,fouc2016-m1_evolve_n}.
The ray tracing method we present in this paper captures the dominant effects
of elastic scattering, while leaving the physics of inelastic scattering and
pair processes for later work.

\begin{table}
  \caption{
    We analyze neutrino interaction processes in terms of these categories.
    $\nu$ without a label represents a neutrino or antineutrino of any flavor,
    $N$ represents a nucleon $n$ or $p$,
    ${}^ZA$ represents a nucleus with mass number $A$ and charge $Z$, and
    $\gamma$ represents a high-energy photon.
    A prime indicates a change in that particle's energy.
  }
  \label{tab:neutrino_processes}
  \begin{tabularx}{\columnwidth}{X X}
    \hline \hline
    absorption/emission
    & $\nu_e + n \leftrightarrow e^- + p$                          \\
    & $\bar{\nu}_e + p \leftrightarrow e^+ + n$                    \\
    & $\nu_e + {}^AZ \leftrightarrow e^- + {}^A(Z+1)$              \\
    \hline
    elastic scattering
    & $\nu + N \leftrightarrow \nu + N$                            \\
    & $\nu + {}^AZ \leftrightarrow \nu + {}^AZ$                    \\
    \hline
    inelastic scattering
    & $\nu + e^- \leftrightarrow \nu' + e^{-'}$                    \\
    & $\nu + e^+ \leftrightarrow \nu' + e^{+'}$                    \\
    \hline
    thermal pair processes
    & $\nu + \bar{\nu} \leftrightarrow e^{-} + e^{+}$              \\
    & $\nu + \bar{\nu} + N + N \leftrightarrow N' + N'$            \\
    & $\nu + \bar{\nu} \leftrightarrow \gamma$                     \\
    \hline \hline
  \end{tabularx}
\end{table}

Ray tracing is ideally suited to problems requiring detailed knowledge of
radiation distribution functions over small regions of spacetime:
for example along a matter or radiation trajectory,
or over a small volume outside a source.
Furthermore, our method is time-dependent, allowing us to compute radiation
fields in dynamical systems. But it is formulated as a post-processing step,
with the ray tracing computated on volume data saved in several steps
of the fluid evolution.
Thus for dynamical systems the memory demands can be prohibitively large.

More fundamentally, though, ray tracing is limited by its naive
treatment of the Boltzmann Equation (Eqn.~\ref{eqn:boltzmann}),
a treatment which essentially decouples different neutrino momenta and species.
When solving for $f^{\nu_\sigma}(x^\alpha;p_\beta)$,
we have no information about the distribution at different momenta
$f^{\nu_\sigma}(x^\alpha;p'_\beta)$,
or about the distribution of the relevant antineutrino species
$f^{\bar{\nu}_\sigma}(x^\alpha;p'_\beta)$;
these missing data are essential ingredients
for the source terms of the Boltzmann Equation describing the creation and
destruction of neutrinos due to the interaction processes described in
Tab.~\ref{tab:neutrino_processes}.

In this paper we outwit this limitation by incorporating coupled source terms
that depend on either previously-evolved or analytical
estimates of the neutrino fields.
To incorporate scattering processes in our method, we employ estimates of the
lowest-order moment of the neutrino distribution function
computed in an M1 transport simulation.
If moments are not available either from
an M1 transport evolution or a trustworthy analytical estimate,
we may drop the coupling terms and our method reduces to the current
state-of-the-art ray tracing methods neglecting scattering.
Our method is not a standalone radiation transport scheme,
but serves as the final component of a hybrid scheme,
piggy-backing on a lower-order radiation transport method as a
post-processing step.

In Sec.~\ref{sec:formulation} we derive the ray tracing equations from the
Boltzmann Equation and describe our numerical scheme.
In Sec.~\ref{sec:tests} we present tests of the code.
In Sec.~\ref{sec:applications} we present neutrino fields in
the dynamical environment following the merger of two neutron stars,
and compute neutrino flavor oscillation along an outgoing ray,
including the effects of coherent forward scattering with ambient neutrinos.
In Sec.~\ref{sec:conclusions} we summarize our work and anticipate improvements.

Greek tensor indices ($\alpha, \beta, ...$) range over all four coordinates,
whereas Latin indices ($i, j, ...$) range over the spatial coordinates 1--3,
or over a more general set, e.g. the set of all elastic scattering interactions.
We use naturalized units in which $\{\hbar,c,k_{\rm B}\}=1$.
And for most of the remainder of this article we suppress neutrino species label
$\nu_\sigma$ since the formulation is general to any species.
Where we do reference particular species we use the three categories
relevant to the energy scales of mergers, $\nu_e$, $\bar{\nu}_e$, and
$\nu_x=\left\{\nu_\mu,\bar{\nu}_\mu,\nu_\tau,\bar{\nu}_\tau\right\}$.

\section{Ray tracing formulation}
\label{sec:formulation}
The neutrino distribution function, $f(x^\alpha; p_\beta)$ is an invariant
quantity counting the number of neutrinos in a given six-volume of phase
space centered on $(x^\alpha,p_\beta)$.
The phase space volume elements are defined with respect to a fiducial
observer passing through event $x^\alpha$ with velocity $u^\alpha$:
\begin{align}
  \label{eqn:dV}
  dV & \equiv \sqrt{-\psi} \, dx \, dy \, dz \, u^t \\
  \label{eqn:dP}
  dP & \equiv \frac{1}{\sqrt{-\psi}} \, dp_x \, dp_y \, dp_z \,
  \frac{\varepsilon}{p^t},
\end{align}
where $\psi$ represents the determinant of the spacetime metric,
the index $t$ indicates the time-component of the given four-vector, and
\begin{equation}
  \label{eqn:varepsilon}
  \varepsilon \equiv -p_\mu u^\mu
\end{equation}
is the neutrino energy measured by our observer.
The number of particles in a given six-volume is
\begin{equation}
  dN=\frac{g}{(2\pi)^3}\,f\,dV\,dP,
\end{equation}
where $g$ counts the number of spin states accessible to the
particles ($g=1$ for neutrinos), and $f$ is the distribution function.
Each of $dV$, $dP$, $dN$, and $f$ are spacetime invariants
\citep{debb2009-gr_boltzmann_1, debb2009-gr_boltzmann_2, lind1966-gr_boltzmann}.

We may decompose the neutrino momentum like
\begin{equation}
  \label{eqn:def_momentum}
  p_\beta = \varepsilon (u_\beta + \ell_\beta),
\end{equation}
with $\ell_\beta$ the direction normal subject to the constraints
$u^\alpha \ell_\alpha = 0$ and $\ell^\alpha \ell_\alpha=1$.
With this decompositon we can write the arguments to the distribution function
$f(x^\alpha;\varepsilon,\ell_\beta)$.

Because $\ell_\beta$ is subject to two constraints
(normalization and orthogonality to the observer's velocity)
it has only two remaining degrees of freedom; we make this explicit by defining its
spatial cartesian components with respect to spherical polar angles
\begin{equation}
  \label{eqn:def_direction}
  \ell_\alpha \rightarrow
  q (s,\sin a \cos b,\sin a\sin b,\cos a),
\end{equation}
with $q$ and $s$ functions of $a$ and $b$.
Now our symbol for the distribution function,
$f(x^\alpha;\varepsilon,a,b)$,
makes manifest its seven independent arguments.

We also define a rotated frame
\begin{equation}
  \ell_{\alpha'}=\frac{\partial x^\alpha}{\partial x^{\alpha'}}\,\ell_\alpha \nonumber
\end{equation}
in which the cartesian components of the direction 1-form are defined
\begin{equation}
  \label{eqn:def_direction_primed}
  \ell_{\alpha'} \rightarrow
  q (s,\sin A \cos B,\sin A\sin B,\cos A),
\end{equation}
so that momenta with vanishing polar angle $A$ move outward along coordinate
radial rays.
This transformation is chosen so that for an observer far from the source,
incoming radiation will be concentrated into a narrow beam around $\cos A\approx1$,
independent of that observer's position in coordinate space.
For explicit definitions see App.~\ref{sec:definitions}.

In Minkowski spacetime, and for a stationary observer
$u_\beta\rightarrow(-1,0,0,0)$, we would have $s=0$, $q=1$,
and $\varepsilon=-p_t$. In that case $\cos A$ may be identified
with the familiar forward direction cosine $\mu$
\citep{shu1991-book_radiation,miha1999-foundations}.

\subsection{Boltzmann Equation}
\label{ssec:boltzmann}
In the limit of large oscillation lengths (see Sec.~\ref{ssec:V_nunu}),
neutrino radiation obeys the relativistic Boltzmann Equation,
which, suppressing the arguments of $f$ for simplicity, is written,
\begin{equation}
  \label{eqn:boltzmann}
  \frac{d}{d\lambda}f = C[f],
\end{equation}
where $d/d\lambda$ denotes a derivative with respect to the affine
parameter defining the neutrino momentum (Eqn.~\ref{eqn:geodesic_x} below)
and $C[f]$ is the source term arising from interactions with the medium.
As we will show below in Eqn.~\ref{eqn:proper_length},
the affine parameter has dimension ${\rm length}\,{\rm energy}^{-1}$,
so the source term has dimension ${\rm energy}\,{\rm length}^{-1}$.
The source term varies over phase space $(x^\alpha,p_\beta)$,
and depends locally on the distribution function $f$
and nonlocally on the distribution function of this neutrino and its
antiparticle at different momenta, $f'$ and $\bar{f}'$.
For simplicity we symbolize all of these dependencies with the shorthand $C[f]$.
The various neutrino interactions contributing to $C[f]$ are detailed in
App.~\ref{sec:source_terms}.

We make the right hand side of Eqn.~\ref{eqn:boltzmann} explicit by writing
the source term linear in $f$:
\begin{align}
  \label{eqn:boltzmann_linear}
  \frac{d}{d\lambda}f &=
  \mathscr{E} - \mathscr{K} f \\
  \label{eqn:boltzmann_linear_s}
  &= \mathscr{K}(\mathscr{S}-f)
\end{align}
where we have introduced
$\mathscr{E}$, the invariant total emissivity, and
$\mathscr{K}$, the invariant total opacity.
These describe respectively the energy gained and the energy lost
per length traveled by the neutrino, true scalar quantities
which take identical values for all observers.
In the second form we have introduced the source function
$\mathscr{S}\equiv\mathscr{E}/\mathscr{K}$, which makes manifest the
behavior of the right hand side, driving $f$ toward $\mathscr{S}$
over a lengthscale $\mathscr{K}^{-1}$ in the affine parameter;
or from Eqn.~\ref{eqn:proper_length} below, over a proper lengthscale
$\varepsilon/\mathscr{K}$ measured by the fiducial observer.

These coefficients are computed by considering their dependence
on neutrino and antineutrino distribution functions at other momenta
(i.e.\ Fermi-blocking).
We consider the two dominant classes of interactions in this work:
the absorption/emission (AE) and elastic scattering (SE) processes
listed in Tab.~\ref{tab:neutrino_processes};
note that we also include the important thermal pair processes (PP) for
heavy-lepton neutrinos by incorporating an approximate pair emissivity
into their absorption/emission coefficients;
see App.~\ref{sec:source_terms} for details.
Thus we separate these coefficients:
\begin{align}
  \label{eqn:inv_emissivity_components}
  \mathscr{E} &= \mathscr{E}_{\rm AE} + \mathscr{E}_{\rm SE},\\
  \label{eqn:inv_opacity_components}
  \mathscr{K} &= \mathscr{K}_{\rm AE} + \mathscr{K}_{\rm SE}.
\end{align}

The absorption/emission coefficients are computed from sums
over the relevant emissivities and opacities
for the reactions
\begin{align}
  \nu_e + n            &\leftrightarrow e^- + p,\nonumber\\
  \bar{\nu}_e + p      &\leftrightarrow e^+ + n,\nonumber\\
  \nu_e + {}^AZ        &\leftrightarrow e^- + {}^A(Z+1),\nonumber
\end{align}
with ${}^AZ$ representing a nucleus of mass number $A$ and charge $Z$.
In terms of the emissivity $j(\varepsilon)$ describing number of neutrinos
of energy $\varepsilon$ emitted per length,
and the absorption opacity $\chi_a(\varepsilon)$
describing the number absorbed per length,
the coefficients are
\begin{align}
  \label{eqn:ae_emissivity_summed}
  \mathscr{E}_{\rm AE}(\varepsilon)
  &= \varepsilon \sum_{i\,{\rm reactions}} j_i(\varepsilon), \\
  \label{eqn:ae_opacity_summed}
  \mathscr{K}_{\rm AE}(\varepsilon)
  &= \frac{1}{1-f^{\rm eq}(\varepsilon)} \,
  \varepsilon \sum_{i\,{\rm reactions}} \chi_{{\rm a},i}(\varepsilon),
\end{align}
where $f^{\rm eq}$ is the distribution function of neutrinos in radiative
equilibrium with the matter, i.e. the Fermi-Dirac distribution function
\begin{equation}
  \label{eqn:feq}
  f^{\rm eq}(\varepsilon) \equiv
  \left(1+e^{\varepsilon/(k_{\rm B} T) -\eta_\nu}\right)^{-1},
\end{equation}
with the neutrino chemical potentials dependent on the local density,
temperature, and composition of the fluid via the neutron, proton,
and electron chemical potentials:
$\eta_{\nu_e}=-\eta_{\bar{\nu}_e}=\eta_p-\eta_n+\eta_{e^-}$ and
$\eta_{\nu_x}=0$.
The appearance of $f^{\rm eq}$ in Eqn.~\ref{eqn:ae_opacity_summed} is due to
the Fermionic nature of the neutrinos, causing $\mathscr{K}_{\rm AE}$ to be
different than the simple absorption opacity,
a phenomenon called stimulated absorption \cite{burr2006-neutrino_opacities}.
By detailed balance of the absorption/emission reactions, we may alternatively
write the emissivity in terms of the equilibrium distribution function:
\begin{equation}
  \label{eqn:ae_emissivity_feq}
  \mathscr{E}_{\rm AE}(\varepsilon)
  = \mathscr{K}_{\rm AE}(\varepsilon) f^{\rm eq}(\varepsilon).
\end{equation}
Note that the stimulated absorption coefficient $\mathscr{K}_{\rm AE}$ is
identical to the coefficient $\kappa^*$
defined in \cite{hari2010-gr_nunubar_collapsar}.
See App.~\ref{ssec:sources_ae} for details.

The elastic scattering coefficients are computed from
a background field, and a sum over opacities for
the reactions
\begin{align}
  \nu + N     &\leftrightarrow \nu + N,\nonumber\\
  \nu + {}^AZ &\leftrightarrow \nu + {}^AZ,\nonumber
\end{align}
with $N$ standing in for either $n$ or $p$.
In terms of a background field $\Phi(\varepsilon)$ describing
the number of neutrinos of energy $\varepsilon$ present at this event,
and the scattering opacity $\chi(\varepsilon)$ describing the
number scattered to other directions per length,
the coefficients are
\begin{align}
  \label{eqn:se_emissivity}
  \mathscr{E}_{\rm SE}(\varepsilon)
  &= \mathscr{K}_{\rm SE}(\varepsilon) \, \Phi(\varepsilon) \, \\
  \label{eqn:se_opacity_summed}
  \mathscr{K}_{\rm SE}(\varepsilon)
  &= \varepsilon \sum_{i\,{\rm reactions}} \chi_{{\rm s},i}(\varepsilon).
\end{align}
In the isotropic limit of trapped radiation,
$\Phi$ is equivalent to $f^{\rm eq}$;
in the free-streaming limit at a distance $r$ from a source,
$\Phi$ attenuates as $r^{-2}$.
See App.~\ref{ssec:sources_se} for details.

With these definitions we write a separated form of
Eqn.~\ref{eqn:boltzmann_linear_s}:
\begin{equation}
  \label{eqn:boltzmann_split}
  \frac{d}{d\lambda}f =
  \mathscr{K}_{\rm AE}(f^{\rm eq}-f)
  + \mathscr{K}_{\rm SE}(\Phi-f)
\end{equation}
From Eqn.~\ref{eqn:boltzmann_split} we see that
absorption/emission interactions drive the distribution function
toward $f^{\rm eq}$ over an affine lengthscale $\mathscr{K}_{\rm AE}^{-1}$,
and elastic scattering interactions drive it
toward $\Phi$ over an affine lengthscale $\mathscr{K}_{\rm SE}^{-1}$.
(Eqn.~\ref{eqn:proper_length} translates affine length to proper length
for a given observer; in this case the proper length scale is
$\varepsilon/\mathscr{K}$.)


This paper introduces the use of neutrino densities and fluxes evolved in an
M1 transport simulation to estimate the background field, $\Phi$.
App.~\ref{ssec:sources_se} details a method to calculate
$\Phi(\varepsilon)$ in two different ways:
\begin{itemize}
\item
  the \emph{spectral} method using densities and fluxes
  extracted from a simulation evolved over multiple energy groups
  to compute the background field with Eqn.~\ref{eqn:background_phi},
\item
  the \emph{gray} method using energy-integrated densities and fluxes
  extracted from a gray simulation and approximating the
  energy distribution with Eqns.~\ref{eqn:J_from_gray} and
  \ref{eqn:H_from_gray}.
\end{itemize}

\subsection{Trajectories}
\label{ssec:trajectories}
Each trajectory is uniquely labeled by a pair of vectors
giving an event on the trajectory, $x^\alpha$,
and the momentum at that event, $p_\beta$.
To designate a family of intersecting trajectories, we keep constant either
the emission event $x^\alpha_e$ or
the observation event $x^\alpha_o$.

Neutrino trajectories obey the geodesic equation, which may be decomposed into
the coupled first-order equations
\begin{equation}
\label{eqn:geodesic_x}
  \frac{d x^\alpha}{d\lambda} = p^\alpha,
\end{equation}
and
\begin{equation}
\label{eqn:geodesic_p}
  \frac{d p_\beta}{d\lambda} = -\Gamma^\alpha_{\beta\gamma} p^\gamma p_\alpha,
\end{equation}
where $p^\alpha=\psi^{\alpha\beta}p_\beta$,
$\psi^{\alpha\beta}$ is the inverse of the spacetime metric $\psi_{\alpha\beta}$,
and $\Gamma^\alpha_{\beta\gamma}$ are the standard connection coefficients,
\begin{equation}
  \label{eqn:christoffel}
  \Gamma^\alpha_{\beta\gamma} =
  \frac{1}{2} \psi^{\alpha\mu}
  (\psi_{\mu\beta,\gamma} + \psi_{\mu\gamma,\beta} - \psi_{\beta\gamma,\mu}),
\end{equation}
with the comma denoting a partial derivative
$\psi_{\mu\beta,\gamma}=\partial_\gamma\,\psi_{\mu\beta}$.

\begin{figure}
  \includegraphics[width=\columnwidth]{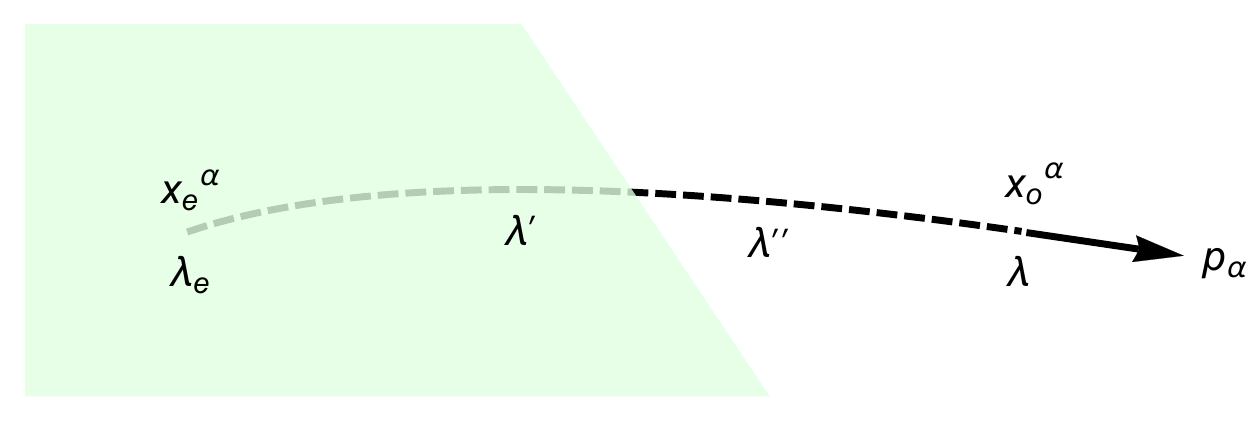}
  \caption{Affine parameterization of a neutrino trajectory of momentum
    $p_\alpha$.
    The fiducial observer with velocity $u^\alpha$ sits at $x^\alpha_o$,
    the neutrino emission event is at $x^\alpha_e$. The affine parameter
    increases from the emission event: $\lambda_e<\lambda'<\lambda''$.
    The green slab represents dense matter.}
  \label{fig:affine_param}
\end{figure}

Each trajectory is parameterized by affine parameter, $\lambda$, increasing in
the direction of $\ell_\beta$. We label $\lambda=\lambda_e$ at $x^\alpha_e$,
as in Fig.~\ref{fig:affine_param}.
If we multiply Eqn.~\ref{eqn:geodesic_x} by $u_\alpha \equiv dx_\alpha/ds$,
we find the element of proper distance traversed by the
neutrino as measured by the fiducial observer $u^\alpha$ is
\begin{equation}
  \label{eqn:proper_length}
  ds=\varepsilon \,d\lambda.
\end{equation}

\subsection{The Formal Solution}
\label{ssec:rendering_eqn}
We can integrate Eqn.~\ref{eqn:boltzmann_split} directly, 
with the solution split into a boundary, absorption/emission, and
a scattering term, $f=f_{\rm bdry}+f_{\rm AE}+f_{\rm SE}$:
\begin{align}
  \label{eqn:rendering_fbdry}
  f_{\rm bdry}(\lambda,\lambda_e)
  &= f(\lambda_e) e^{-\tau(\lambda,\lambda_e)}, \\
  \label{eqn:rendering_fae}
  f_{\rm AE}(\lambda,\lambda_e)
  &= \int_{\lambda_e}^\lambda d\lambda'\,e^{-\tau(\lambda,\lambda')}
  \mathscr{K}_{\rm AE}(\lambda')f^{\rm eq}(\lambda'), \\
  \label{eqn:rendering_fse}
  f_{\rm SE}(\lambda,\lambda_e)
  &= \int_{\lambda_e}^\lambda d\lambda'\,e^{-\tau(\lambda,\lambda')}
  \mathscr{K}_{\rm SE}(\lambda')\Phi(\lambda'),
\end{align}
where the optical depth is defined,
\begin{equation}
  \label{eqn:optical_depth}
  \tau(\lambda,\lambda') \equiv \int_{\lambda'}^\lambda
  d\lambda'' \, \mathscr{K}(\lambda''),
\end{equation}
and the parameterization conventions are depicted in
Fig.~\ref{fig:affine_param}.
Note that Eqn.~\ref{eqn:optical_depth} employs the total absorption
plus scattering opacity, so that the optical depth attenuating the
integrands of Eqns.~\ref{eqn:rendering_fae} and \ref{eqn:rendering_fse}
is the total optical depth.


\subsection{Moments of the Distribution Function}
\label{ssec:moments}
We may take angular moments of the distribution function:
\begin{align}
  \label{eqn:J}
  J(\varepsilon) &=
  \frac{\varepsilon^3}{(2\pi)^3} \oint d\Omega' f(\varepsilon, \ell'_\beta) \\
  \label{eqn:Ha}
  H^\mu(\varepsilon) &=
  \frac{\varepsilon^3}{(2\pi)^3} \oint d\Omega' f(\varepsilon, \ell'_\beta) \ell'^\mu \\
  \label{eqn:Sab}
  S^{\mu\gamma}(\varepsilon) &=
  \frac{\varepsilon^3}{(2\pi)^3} \oint d\Omega' f(\varepsilon, \ell'_\beta) \ell'^\mu \ell'^\gamma,
\end{align}
defining the specific energy density,
specific momentum density, and
specific radiation pressure tensor, respectively.
Here ``specific'' refers to the quantity being integrable over neutrino energy.
Integrals~\ref{eqn:J}--\ref{eqn:Sab} are performed over a solid angle in
momentum space while holding $\varepsilon$ constant:
$d\Omega \equiv d(\cos a)\,db$.
We also make use of the specific number density and specific number flux defined
\begin{align}
  \label{eqn:G}
  G(\varepsilon) &=
  \frac{\varepsilon^2}{(2\pi)^3} \oint d\Omega' f(\varepsilon, \ell'_\beta), \\
  \label{eqn:Ka}
  K^\mu(\varepsilon) &=
  \frac{\varepsilon^2}{(2\pi)^3} \oint d\Omega' f(\varepsilon, \ell'_\beta) \ell'^\mu.
\end{align}

The energy-integrated moments take the form
\begin{equation}
  \label{eqn:J_H_S_eps_integrated}
  X = \int_0^\infty d\varepsilon \, X(\varepsilon),
\end{equation}
with $X$ standing in for any of $\{J,H^\mu,S^{\mu\gamma}\}$ or
$\{G,K^\mu\}$,
the first having dimension ${\rm energy}\,{\rm length}^{-3}$,
and the second ${\rm length}^{-3}$.

We compute moments for a particular observer
by specifying the four-velocity in Eqn.~\ref{eqn:def_momentum}.
Two choices are particularly useful:
an observer stationary in the coordinate frame (i.e. Eulerian),
or one stationary in the fluid frame (i.e. comoving)
\cite{smar1980-gr_hydro}.
Explicit definitions are given in App.~\ref{sec:definitions}.
We distinguish moments computed for an Eulerian observer with a tilde,
e.g. $\tilde{J}$, $\tilde{H}^\mu$, $\tilde{S}^{\mu\nu}$;
note that these three Eulerian moments are identical to the lab-frame moments
$E$, $F^\mu$, and $P^{\mu\nu}$
defined in
\cite{shib2011-truncated_moment, ocon2015-gr1d_with_nu, fouc2015-m1_nsbh}.

\subsection{Numerical Implementation}
\label{ssec:numerical}
Much of our numerical implementation is borrowed from the geodesic evolution
system described in~\cite{bohn2016-code}.
We integrate Eqns.~\ref{eqn:geodesic_x} and~\ref{eqn:geodesic_p} in the form
given by \citep{hugh1994-eh_finding},
and Eqns.~\ref{eqn:rendering_fae}, \ref{eqn:rendering_fse},
and \ref{eqn:optical_depth} in the form given below.
By using the time-component of Eqn.~\ref{eqn:geodesic_x} ($dt=d\lambda \, p^t$)
we may transform the integrations to coordinate time.
The coupled system of ordinary differential equations is
\begin{align}
  \label{eqn:num_x}
  \frac{dx^i}{dt} &=
  g^{ij} \frac{p_j}{p^t} - \beta^{i},\\
  \label{eqn:num_p}
  \frac{dp_i}{dt} &=
  -\alpha \alpha_{,i} p^t
  + \beta^k_{,i} p_k
  - \frac{1}{2} g^{jk}_{,i}\frac{p_j p_k}{p^t},\\
  \label{eqn:num_tau}
  \frac{d\tau}{dt} &=
  - \frac{1}{p^t}\mathscr{K},\\
  \label{eqn:num_fae}
  \frac{df_{\rm AE}}{dt} &=
  \frac{1}{p^t}e^{-\tau}\mathscr{K_{\rm AE}} \,f^{\rm eq},\\
  \label{eqn:num_fse}
  \frac{df_{\rm SE}}{dt} &=
  \frac{1}{p^t}e^{-\tau}\mathscr{K_{\rm SE}} \,\Phi.
\end{align}

We integrate each ray until we reach a terminal optical depth of
$\tau_{\rm term}$ at the earliest effective emission event $x_e^\alpha$.
The concept of earliest emission event is a fictitious construct we use
to allow us to truncate the integration at an event along the ray where any
further additions to the field are negligible due to the large optical depth
between $x_e^\alpha$ and $x_o^\alpha$.
We choose $\tau_{\rm term}=14$ so that $e^{-\tau_{\rm term}}<10^{-6}$,
and we then discard the contribution of $f_{\rm bdry}$
(Eqn.~\ref{eqn:rendering_fbdry}).

Since we don't know the emission event a priori,
we follow the integration backwards in time,
from $t_o$ to $t_e$.
We begin each integration by setting initial values for the variables at $t_o$:
the observer specifies $x^i_o$ and $p_{i,o}$,
and we set $f_{{\rm AE},o}=f_{{\rm SE},o}=0$ and $\tau_o=0$.

We integrate Eqns.~\ref{eqn:num_x}--\ref{eqn:num_fse} with adaptive step sizes,
using the 3rd order Runge-Kutta algorithm
which produces an error estimate by comparing the 3rd and 2nd order solutions.
After each step is taken, the errors for each of the nine variables of
Eqns.~\ref{eqn:num_x}--\ref{eqn:num_fse} are compared to an absolute and
a relative threshold.
If the error in any variable exceeds its threshold
the step size is decreased and the step recomputed;
if all errors are below threshold, the next step size is increased.
In practice, the controlling errors come from the radiation variables
$\tau$, $f_{\rm AE}$, and $f_{\rm SE}$,
for which the relative tolerances are set to $6\times10^{-4}$,
and the absolute tolerances are set to $6\times10^{-24}$.

We integrate these equations through the simulated spacetime
over which the following volume data are known:
the spacetime metric $\psi_{\alpha\beta}$,
its derivatives $\psi_{\alpha\beta,\gamma}$,
the fluid velocity $u_i$,
Lorentz factor $W$,
density $\varrho$,
temperature $T$,
and electron fraction $Y_e$,
all defined in App.~\ref{sec:definitions}.
These fields are computed in a preprocessing step before ray tracing
and stored in spectral representation.
If they are computed from a hydrodynamical simulation, they may be saved to disk
at either one or several specified coordinate times and interpolated
with spectral interpolation in space,
and 1st-order polynomial interpolation in time
(as described in \cite[App.~B]{bohn2016-code}).
If computed from a stationary solution to the
general relativistic hydrodynamics equations no time interpolation is needed.
In this paper for simplicity and to limit computational memory loads,
we use only stationary analytical solutions or
quasi-stationary configurations evolved in simulation,
thus using one time slice and no time interpolation in every case.

\section{Code Tests}
\label{sec:tests}

To test the algorithm, we integrate the ray tracing equations
(Eqns.~\ref{eqn:num_x}--\ref{eqn:num_fse}) for various observers
in the following configurations. This suite of configurations
defines a hierarchy of increasing physical realism:
beginning with a homogeneous medium of effectively
infinite extent (i.e.\ optically thick) and progressing to a model of a
1D pre-supernova, post-bounce collapse profile
evolved using an M1 transport hydrodynamical simulation.
In the pre-supernova model, we compare ray tracing distributions
to those calculated in a Monte Carlo transport simulation.

In the following we present two forms of the integrated distribution functions,
Eqns.~\ref{eqn:num_fae} and \ref{eqn:num_fse}:
\begin{itemize}
\item the \emph{scat} form including elastic scattering,
  for which the solution is
  $f=f_{\rm AE,scat}+f_{\rm SE,scat}$,
\item the \emph{noscat} form treating only absorption/emission interactions
  by setting $\mathscr{K}_{\rm SE}=0$,
  for which the solution is
  $f=f_{\rm AE,noscat}$.
\end{itemize}
Note that $f_{\rm AE,scat} \neq f_{\rm AE,noscat}$ because $f_{\rm AE}$
is integrated using the absorption plus scattering optical depth in
the first form and using only the absorption optical depth in the second form.
So there is no simple algebraic relation between the \emph{scat} and
\emph{noscat} distribution functions.
The differences between the two methods is apparent
in Fig.~\ref{fig:cumulative_f_homogeneous} below.
Before this work, the \emph{noscat} form of the equations was the standard
for ray tracing, though many authors included the scattering optical depth in
the integration of Eqn.~\ref{eqn:num_fae}
(see for example \cite{hari2010-gr_nunubar_collapsar}).

\subsection{Infinite Homogeneous Slab:
  Testing Thermodynamic Equilibrium}
\label{ssec:test_equilibrium}
In an optically thick region the radiation field is in thermodynamic equilibrium
with the matter.
We set up a large slab of matter in a Minkowski spacetime
representative of the fluid thermodynamic state at a radius of 50~km
in the test presented in Sec.~\ref{ssec:test_collapse}---a
massive collapsed star following core bounce---where
$\varrho=10^{11}\,{\rm g}\,{\rm cm}^{-3}$, $T=3.7$~MeV, and $Y_e=0.12$.
For comparision with that test we use the LS180 equation of state
\cite{latt1991-nuc_eos, ocon2010-gr1d}
in which the equilibrium $\nu_e$ neutrino chemical potential is
$\eta_{\nu_e}=-0.1555$
(with $\eta_{\bar{\nu}_e}=-\eta_{\nu_e}$ and $\eta_{\nu_x}=0$).

The opacity table is computed using \lstinline{NuLib}
\cite{ocon2015-gr1d_with_nu}
and is identical to the LS180 opacity table used in the referenced paper.
The scattering opacity is computed taking into account elastic scattering on
nucleons, alpha particles, and heavy nuclei.
The absorption opacities consist of electron neutrino absorption on neutrons
and heavy nuclei as well as electron antineutrino absorption on protons.
We use Kirchhoff's law to compute emissivities based on these absorption
opacities.
For heavy lepton neutrinos we consider thermal emission processes including
electron-positron annihilation and nucleon-nucleon Bremsstrahlung.
The table is stored on a grid covering energy, density, temperature,
and composition
ranges spanning $\varepsilon_i\in[1,280.5]$~MeV logarithmically,
$\varrho\in[10^6,6.3\times10^{15}]\,{\rm g}\,{\rm cm}^{-3}$ logarithmically,
$T\in[0.05,200]$~MeV logarithmically, and 
$Y_e\in[0.035,0.55]$ linearly,
with grid extents$\{18,82,65,51\}$ respectively.
Interpolation is performed linearly.

We use the equilibrium distribution functions to define background
fields for the \emph{scat} case:
in the \emph{spectral} method we use
$J(\varepsilon)=\varepsilon^3 f^{\rm eq}(\eta_\nu,T;\varepsilon)/(2\pi^2)$,
and in the \emph{gray} method we use
$J=C T^4 \mathscr{F}_3(\eta_\nu)/(2\pi^2)$ and
$G=C T^3 \mathscr{F}_2(\eta_\nu)/(2\pi^2)$,
using the Fermi integrals defined in Eqn.~\ref{eqn:fermi_integral},
and with $C$ the conversion constant from ${\rm energy}^3$ to ${\rm length}^{-3}$.

\begin{figure}
  \resizebox{\columnwidth}{!}{\input{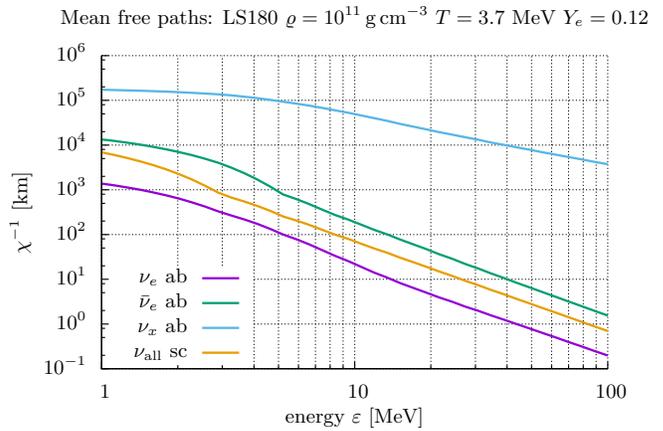}}
  \caption{Mean free paths
    representative of the post-bounce collapse profile (presented in
    Sec.~\ref{ssec:test_collapse}) at
    50~km, where $\varrho=10^{11}\,{\rm g}\,{\rm cm}^{-3}$,
    $T=3.7$~MeV, and $Y_e=0.12$; this thermodynamic state is used in
    the tests with homogeneous matter distributions
    (presented in Secs.~\ref{ssec:test_equilibrium}--\ref{ssec:test_ab_star}).
    In the fluid rest frame the mean free paths are given by the inverses of the opacities
    $\chi^*_a=\mathscr{K}_{\rm AE}/\varepsilon$ for absorption and
    $\chi_s=\mathscr{K}_{\rm SE}/\varepsilon$ for elastic scattering interactions
    (see App.~\ref{sec:source_terms} for definitions).
    Note that elastic scattering opacities are identical across species
    below energies at which weak magnetism plays a role \cite{horo2002-weak_mag}.
    In these data they are exactly identical
    because we have turned off weak magnetism in our opacity calculations
    in order to compare our results with the historical literature.
    Computed using \lstinline{NuLib}.
  }
  \label{fig:mfps-50km}
\end{figure}

Fig.~\ref{fig:mfps-50km} presents the neutrino mean free paths
at this thermodynamic point over two energy decades.
For this test we choose a slab large enough for neutrinos of all
energies to be trapped.
We sample the distribution function with ray tracing over a uniform grid
of 40 points in energy $\varepsilon \in (0,100)$~MeV.
The domain extends to $\tau_{\rm term}\,L_{\rm mfp,max}\approx10^7\,{\rm km}$,
since our ray tracing algorithm integrates rays to terminal optical
depths of $\tau_{\rm term}\geq14$.

\begin{figure}
  \resizebox{\columnwidth}{!}{\input{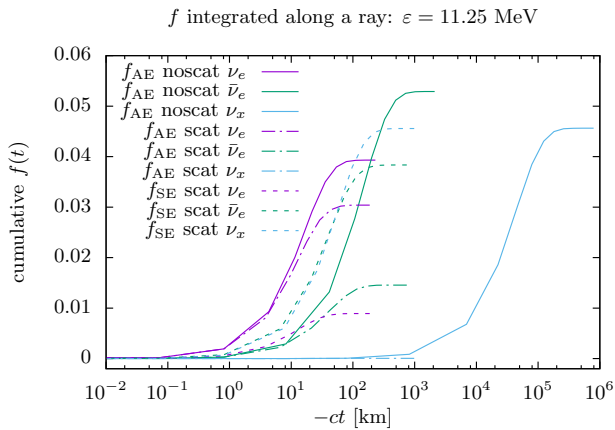}}
  \caption{Cumulative distribution functions at $\varepsilon=11.25$~MeV
    in the homogeneous infinite slab test
    (presented in Sec.~\ref{ssec:test_equilibrium}).
    The integration proceeds from left to right, or backwards in time $t$.
    Each ray terminates when it achieves a total optical depth
    greater than $\tau_{\rm term}=14$.
    The points plotted correspond to the time steps chosen by the adaptive
    time-stepping algorithm described in Sec.~\ref{ssec:numerical}.
  }
  \label{fig:cumulative_f_homogeneous}
\end{figure}

Fig.~\ref{fig:cumulative_f_homogeneous} displays the cumulative distribution
function integrated along the ray, for each of the three species and using
both methods, \emph{noscat} and \emph{scat}, at the single energy
$\varepsilon=11.25$~MeV. We display
Eqns.~\ref{eqn:num_fae} and \ref{eqn:num_fse} in their integral form:
\begin{equation}
  \label{eqn:cumulative_f}
  f(t)
  = - \int_{0}^t \frac{dt'}{\varepsilon}e^{-\tau(0,t')}
  \mathscr{K}(t')f^{\rm eq}(t'),
\end{equation}
and the integration proceeds backwards in time, from $t=0$
to some terminal $t<0$.
The figure shows this backwards-in-time integration proceeding left to right.

In Fig.~\ref{fig:cumulative_f_homogeneous} we see some expected features.
The final distribution functions asymptote to their equilibrium Fermi-Dirac
levels at this energy
$f^{\rm eq}_{\{\nu_e,\bar{\nu}_e,\nu_x\}}=\{0.039, 0.053, 0.046\}$
within a few mean free paths;
in the \emph{scat} cases it is the sums $f_{\rm AE}+f_{\rm SE}$ that
achieve these values.
And the lengths of the rays are proportional to the mean free paths $L$,
which obey the hierarchy $L_{\nu_x}>L_{\bar{\nu}_e}>L_{\nu_e}$;
in the \emph{scat} cases these lengths are less than in the \emph{noscat} cases,
since the total mean free paths are less than the absorption mean free paths
(significantly so for $\nu_x$, negligibly so for $\nu_e$).

\begin{figure}
  \resizebox{\columnwidth}{!}{\input{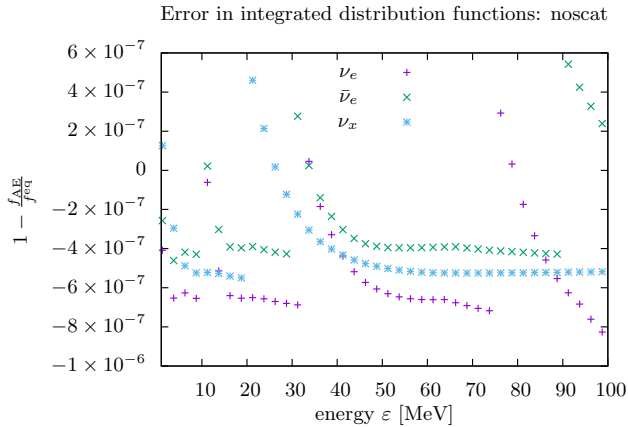}}
  \caption{Relative error in integrated equilibrium distribution functions
    in the infinite homogeneous slab
    (presented in Sec.~\ref{ssec:test_equilibrium}).
    Plotted here are the relative differences between the final
    $f_{\rm AE}$ in the \emph{noscat} case
    (see Fig.~\ref{fig:cumulative_f_homogeneous})
    and the equilibrium distribution functions given by
    Eqn.~\ref{eqn:feq}.
    The source of this error is the discarded boundary term
    $f_{\rm bdry}\sim10^{-6}$, described in Sec.~\ref{ssec:numerical}.
  }
  \label{fig:homogeneous_isotropic}
\end{figure}

As Fig.~\ref{fig:mfps-50km} reveals, integrating the distribution functions
over the energies $\varepsilon\in(0,100)$~MeV
probes the numerical solution over length scales from
0.1~km to $10^{5}\,{\rm km}$.
Fig.~\ref{fig:homogeneous_isotropic} shows the error in our results for
the \emph{noscat} case.
In this simple case, the dominant source of error is the neglected
boundary term with a relative scale of $e^{-\tau_{\rm term}}\approx10^{-6}$.
In the following inhomogeneous configurations,
errors from the integration dominate over the boundary term.
In this test and only this test, we used a higher-order integrator, a 5th
order Dormand-Prince algorithm \cite{pres2007-nr_3rd_ed}. This is because
our default 3rd order Runge-Kutta algorithm estimates a vanishing error
in this configuration.

\subsection{Infinite Homogeneous Moving Slab:
  Testing Doppler Shift}
\label{ssec:test_doppler}
We reproduce the test above, again in Minkowski spacetime,
but with the matter and observer in relative motion.
We use a stationary observer
and fluid moving in the positive $z$-direction:
with $u^\alpha \rightarrow W(1,0,0,v)$ and $W=(1-v^2)^{-1/2}$,
where $W$ is the relativistic Lorentz factor.
All other thermodynamic variables and background fields are unchanged
since our ray integration uses these quantities in the fluid frame.

A stationary observer measures an energy of $\tilde{\varepsilon}$
for a neutrino with momentum
$p_\alpha \rightarrow \tilde{\varepsilon}(-1,\Omega_i)$
and direction $\Omega_i \rightarrow (\sin A, 0, \cos A)$.
In the fluid frame this neutrino has energy $\varepsilon=-u^\alpha p_\alpha$;
therefore the average energy varies with observing angle like
\begin{equation}
  \label{eqn:doppler_avg_eps}
  \langle \tilde{\varepsilon} \rangle(\cos A) =
  \langle \varepsilon \rangle^{\rm eq} \frac{1}{W(1-v \cos A)},
\end{equation}
where the symbol $\langle \tilde{\varepsilon} \rangle(\cos A)$ emphasizes
the functional dependence on $\cos A$.
The equilibrium average energy $\langle \varepsilon \rangle^{\rm eq}$
is given by $T \mathscr{F}_3(\eta_\nu)/\mathscr{F}_2(\eta_\nu)\approx 3T$,
with the Fermi integrals given in Eqn.~\ref{eqn:fermi_integral}.
Eqn.~\ref{eqn:doppler_avg_eps} describes the well-known Doppler effect.

\begin{figure}
  \resizebox{\columnwidth}{!}{\input{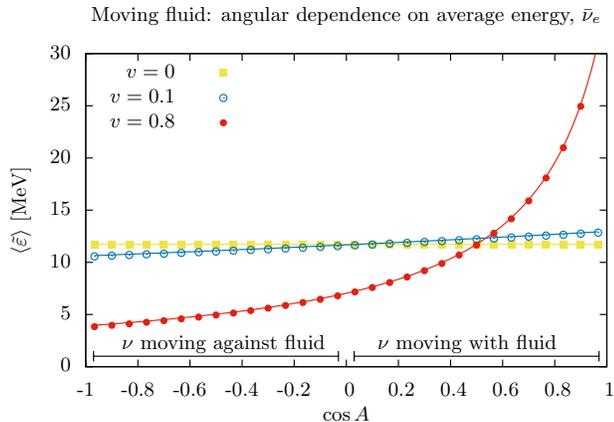}}
  \caption{Average energy of $\bar{\nu}_e$ neutrino fields
    measured by a stationary observer,
    $\langle\tilde{\varepsilon}\rangle(\cos A)$,
    in the moving slab test (presented in Sec.~\ref{ssec:test_doppler}).
    The observer sees neutrinos with $\cos A>0$ to be moving primarily with
    the fluid.
    The points are computed from ray tracing spectra; the lines are the
    analytic formula Eqn.~\ref{eqn:doppler_avg_eps}.
  }
  \label{fig:avg_eps_doppler}
\end{figure}

We sample the distribution function $f(\tilde{\varepsilon},\cos A,B)$
with ray tracing
over a uniform grid of 40 points in energy $\tilde{\varepsilon}\in(0,100)$
and 30 points in angle $\cos A\in(-1,1)$, holding fixed $B=\pi$.
Results are shown in Fig.~\ref{fig:avg_eps_doppler} for
the velocities $v=\{0,0.1,0.8\}$.
The ray tracing results are computed from total densities in each angular bin,
that is
\begin{equation}
  \label{eqn:avg_eps_per_cosA}
  \langle \tilde{\varepsilon} \rangle (\cos A) =
  \frac{\tilde{J}(\cos A)}{\tilde{G}(\cos A)},
\end{equation}
with the Eulerian densities per angular bin given by sums over the samples
\begin{align}
  \label{eqn:G_per_cosA}
  \tilde{G}(\cos A) &=
  \frac{\Delta}{(2\pi)^3} \sum\limits_{m=0}^{N_{\tilde{\varepsilon}}-1}\tilde{\varepsilon}_m^2 f_m(\cos A),\\
  \label{eqn:J_per_cosA}
  \tilde{J}(\cos A) &=
  \frac{\Delta}{(2\pi)^3} \sum\limits_{m=0}^{N_{\tilde{\varepsilon}}-1}\tilde{\varepsilon}_m^3 f_m(\cos A),
\end{align}
with $\Delta \equiv 2\pi \Delta\tilde{\varepsilon}$,
$N_{\tilde{\varepsilon}}$ the number of energy samples,
$f_m(\cos A) \equiv f(\tilde{\varepsilon}_m,\cos A,B)$,
and $m$ labeling each energy bin.
Results are identical for all \emph{scat} and \emph{noscat} methods.
In Fig.~\ref{fig:avg_eps_doppler}
we see the common features of a red-shifted spectrum for receding fluid
($\cos A \sim -1$),
a blue-shifted spectrum for approaching fluid
($\cos A \sim 1$),
and a slightly red-shifted spectrum for fluid moving transverse to the
observer ($\cos A=0$).

\subsection{Idealized Star:
  Testing the Decoupling Regime}
\label{ssec:test_ab_star}
The homogeneous configurations of the previous sections may be extended to
probe the solution outside the optically thick regime
by setting up an idealized homogeneous star, with the thermodynamic variables
constant inside radius $R$ and vanishing outside.
We choose $R=50$~km and place the observer at $r=75$~km,
at which position there is a radiation cone of half-opening angle
$\cos A_{\rm max}\approx0.75$.

The formal solutions of
Eqns.~\ref{eqn:rendering_fae} and \ref{eqn:rendering_fse}
may be directly integrated in this scenario. Assuming Minkowski spacetime
and stationary fluid we have
\begin{align}
  \label{eqn:fae_homogeneous_general}
  f_{\rm AE} &=
  \frac{\chi_a^{*}}{\chi} f^{\rm eq} \left(1-e^{-\chi s}\right), \\
  \label{eqn:fse_homogeneous_general}
  f_{\rm SE} &=
  \frac{\chi_s}{\chi} \Phi \left(1-e^{-\chi s}\right),
\end{align}
where $s$ is the path length traversed by the ray through the star,
\begin{equation}
  \label{eqn:geometric_s}
  s = 2 R \left(1-\frac{r^2}{R^2}(1-\cos^2 A)\right)^{1/2}.
\end{equation}
The total opacity is $\chi=\chi_a^*+\chi_s$, and the stimulated absorption
opacity $\chi_a^*$ and elastic scattering opacity $\chi_s$
are defined in App.~\ref{sec:source_terms}.

Since no analytic form is known for the background field $\Phi$
interior to the star, we examine only the \emph{noscat} case,
with $\chi_s=0$ and $\chi=\chi_a^{*}$.
This scenario has been widely used in the literature as a test for
radiation codes
\cite{smit1997-two_moment,abdi2012-monte_carlo,fouc2015-m1_nsbh}.
We sample the distribution function over a uniform grid of
30 points in angle $\cos A \in (0.734,1)$ (holding fixed $B=\pi$) and
40 points in energy $\tilde{\varepsilon} \in (0,100)$~MeV.
Because of the discontinuity in fluid variables at radius $R$,
we limit the time step size to a maximum of $t_{\rm max}=1.25$~km,
so that as the ray approaches the discontinuity in the homogeneous environment
outside the star, the adaptive time-stepper avoids increasing
the step size beyond the relevant fluid scales.

\begin{figure}
  \resizebox{\columnwidth}{!}{\input{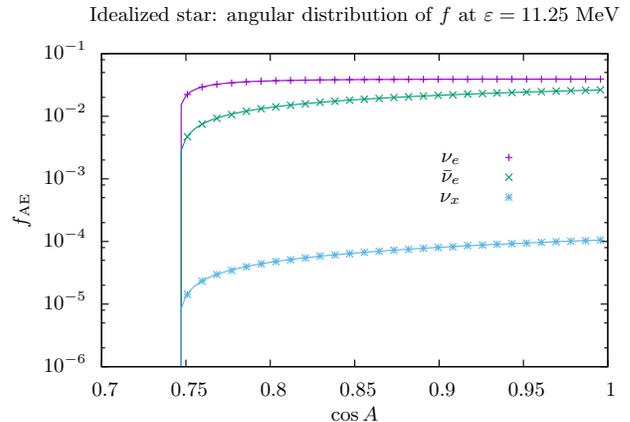}}
  \caption{Distribution functions outside an idealized homogeneous star
    with radius $R=50$~km, and observer at $r=75$~km
    (presented in Sec.~\ref{ssec:test_ab_star}).
    $A$ is the angle between the neutrino momentum and the $\hat{r}$ direction.
    In this plot we display only the samples at energy
    $\varepsilon=11.25\,{\rm MeV}$.
    The points are computed from ray tracing; the lines from the analytic
    solution, Eqn.~\ref{eqn:fae_homogeneous_general}.
  }
  \label{fig:f_absorption_sphere}
\end{figure}

In Fig.~\ref{fig:f_absorption_sphere} we display the samples at
$\varepsilon=11.25$~MeV, along with the analytic functions
specific to each species' equilibrium distribution function and opacity.
As expected only $\nu_e$ saturates at $f^{\rm eq}$, remaining almost
constant across $\cos A$ until we get to rays that pass through a length
of the star comparable to or less than the mean free path at this energy,
16~km. We also see that $\bar{\nu}_e$ comes close to saturating with
a mean free path just over 100~km; and $\nu_x$ is well into the optically
thin regime.

We can explain these features quantitatively by examining the limits
of Eqn.~\ref{eqn:fae_homogeneous_general}, expanding the exponential function
in powers of $\chi s$; the distribution function takes the limiting values
\begin{equation}
  \label{eqn:fae_homogeneous_limits}
  f_{\rm AE}(\cos A) = f^{\rm eq}
  \begin{cases}
    \chi s(\cos A) & s \ll \chi^{-1}\,\textrm{(thin)}\\
    1                  & s \gg \chi^{-1}\,\textrm{(thick)}.
  \end{cases}
\end{equation}
These limits are represented in Fig.~\ref{fig:f_absorption_sphere}:
with $\nu_x$ in the optically thin limit at all viewing angles,
and $\nu_e$ in the optically thick limit at viewing angles
$\cos A \gtrsim 0.8$.

\subsection{Idealized Compact Star:
  Testing Gravitational Redshift and Geodesic Curvature}
\label{ssec:test_gravity}
To test the general relativistic terms in our formulation, which account for
the gravitational redshift and geodesic curvature of the neutrinos,
we sample distribution functions outside an idealized hot compact star,
and compute a neutrino-antineutrino interaction integral describing the
energy deposited per time per volume due to the process
$\nu \bar{\nu} \rightarrow e^{-} e^{+}$.
We describe this code test in detail in
\cite[Sec.~4.3]{deat2015-thesis}, and here give a brief summary.

The $\nu \bar{\nu}$-annihilation integral outside a compact star was computed
semi-analytically in \cite{asan2000-nunubar}. Since then many studies of
$\nu \bar{\nu}$-annihilation in more realistic configurations have used the
compact star as a standard code test \cite{birk2007-nunubar,
  hari2010-gr_nunubar_collapsar,zala2011-nunubar}.
We compute the power density (energy per time per volume) due to
$\nu\bar{\nu}$-annihilation measured by a stationary observer above the star
using:
\begin{equation}
  \label{eqn:q_nunubar}
  q_{\nu\bar{\nu}} = A \int d^3p_{\nu} d^3p_{\bar{\nu}}
  f_{\nu}(p_{\nu j})f_{\bar{\nu}}(p_{\bar{\nu} k})
  \frac{p_{\nu t}+p_{\bar{\nu} t}}{p_{\nu}^t p_{\bar{\nu}}^t}
  (p_{\nu\alpha}p_{\bar{\nu}}^\alpha)^2
\end{equation}
where as in \cite{asan2000-nunubar} we account for the energy redshift to
infinite separation by the energy weighting $p_{\nu t}+p_{\bar{\nu} t}$,
$A=2 c^3 K G_{\rm F}^2$,
the Fermi constant is $G_{\rm F}=5.29\times10^{-44}\,{\rm cm}^2\,{\rm MeV}^{-2}$,
\begin{equation}
  K \left\{ {e \atop \mu\,\tau} \right\} = 
  \frac{1}{6\pi}
  \left(1 \, \left\{ {+ \atop -} \right\}
  \, 4 \sin^2 \theta_w + 8 \sin^4 \theta_w\right)
\end{equation}
and the weak mixing angle is $\sin^2\theta_w=0.23$.

Because Eqn.~\ref{eqn:q_nunubar} has such high dimension,
a simple unigrid integral solution---sampling $f_\nu$ and $f_{\bar{\nu}}$
over fixed step sizes in momentum space---is impractical.
We compute the integral using the adaptive Monte Carlo Vegas technique
\cite{pres2007-nr_3rd_ed},
which iteratively samples those regions of momentum space that contribute
most to the integral. At each iterative stage the algorithm estimates the
error, and terminates when some error threshold is achieved.

In order to stress-test the gravity-dependence of the code,
we choose an unphysically compact star configuration with
radius $R=4.43$~km,
in a Schwarzschild metric with
gravitational radius $R_g=2.95$~km.
To compare to the calculation in \cite{asan2000-nunubar},
instead of integrating the formal solution for $f_\nu$ and $f_{\bar{\nu}}$
using Eqns.~\ref{eqn:rendering_fae} and \ref{eqn:rendering_fse},
we compute only the boundary term using Eqn.~\ref{eqn:rendering_fbdry},
and neglect the attenuation due to the optical depth,
This method is equivalent to transporting the neutrino distribution function
in a state of radiative equilibrium with the matter in the star
up to the observer assuming no interactions along the trajectory.
To define the neutrino distribution function in the star,
we make the star homogeneous, with
temperature $T=5$~MeV and
chemical potentials $\eta_{\nu_e}=-\eta_{\bar{\nu}_e}=0.1$,
and we assume stationary fluid.

The power density deposited by this interaction at a coordinate radius $r=7.38$~km
is computed from formulae in \cite{asan2000-nunubar} as
\begin{equation}
  q_{\nu\bar{\nu}}=6.89\times10^{27}\,{\rm erg}\,{\rm cm^3}\,{\rm s}^{-1}.
\end{equation}
We computed the integral twelve times at an error threshold of 1\% and measured
a mean of
\begin{equation}
  q_{\nu\bar{\nu}}=6.87\pm0.07\times10^{27}\,{\rm erg}\,{\rm cm^3}\,{\rm s}^{-1},
\end{equation}
with the error bars expressing the standard deviation between the twelve
calculations. Each run computed the integral using $N$ samples of the integrand
(requiring $2N$ rays, one for each sample of $f_\nu$ and $f_{\bar{\nu}}$),
with $N$ ranging from 56,000 to 72,000.

The success of this test gives us confidence in the code's ability to handle a
general spacetime metric, since errors in gravitational redshift would have
affected samples of $f$ in the integrand (e.g.\ sampling the distribution function
at the wrong local energy), and errors in geodesic integration would have
affected the angular size of the star (e.g.\ causing the star to look larger
or smaller).

\subsection{Post-Bounce Collapse Profile:
  Testing Scattering}
\label{ssec:test_collapse}
To test our scattering treatments we calculate neutrino fields outside a
collapsed 15~${\rm M}_\odot$ star, 100~ms after core bounce,
comparing ray tracing fields to those from a Monte Carlo transport calculation.
Elastic scattering interior to the shock at $r\approx150$~km
significantly modifies the neutrinos' spectra,
and the extended envelope outside the shock becomes a source of
higher-than-average-energy neutrinos.

The 1D matter profile and M1 transport evolution are computed using
the open source supernova evolution code \lstinline{GR1D}
\footnote{\url{http://www.gr1dcode.org}}
\cite{ocon2010-gr1d, ocon2015-gr1d_with_nu},
using a progenitor profile from \cite{woos1995-sn_progenitors}.
The matter is described by the LS180 equation of state \cite{latt1991-nuc_eos},
and the opacities are computed and stored in a table
as described in Sec.~\ref{ssec:test_equilibrium}.
This standard test is also presented for example in
\cite{ocon2015-gr1d_with_nu,fouc2015-m1_nsbh,abdi2012-monte_carlo}.

The matter profile and background field are stored on a spherical
pseudo-spectral grid composed of 11 spherical-shell subdomains
\cite{kidd2000-spec,szil2009-spec}
comprising a total of 62 radial grid points spaced approximately
logarithmically across $r\in(0,740)$~km.
The background scattering field $\Phi(\varepsilon)$ is supplied by
\lstinline{GR1D} in the form of $J(\tilde{\varepsilon}_i)$,
with $\varepsilon_i$ representing 18 energy groups
identical to those in the \lstinline{NuLib} table described in
Sec.~\ref{ssec:test_equilibrium}.
For the \emph{spectral} method we use $J(\tilde{\varepsilon}_i)$ directly,
using zeroth-order interpolation between energy groups;
for the \emph{gray} method we use $J=\sum J(\tilde{\varepsilon}_i) \Delta\tilde{\varepsilon}_i$,
and $G=\sum J(\tilde{\varepsilon}_i) \Delta\tilde{\varepsilon}_i/\tilde{\varepsilon}_i$ instead,
with $\Delta\tilde{\varepsilon}_i$ the bin-width of the $i$th energy group.

For fiducial neutrino distributions we use the matter profile
as input into a Monte Carlo radiation transport calculation using
open source neutrino transport code \lstinline{Sedonu}
\footnote{\url{https://bitbucket.org/srichers/sedonu}}
\cite{rich2015-monte_carlo}.
To homogenize the physics modeled across these three treatments
(M1 transport to provide the background fields,
ray tracing to compute neutrino distributions, and
Monte Carlo transport for a fiducial comparison)
we turn off inelastic scattering where it is included (in the Monte Carlo code),
and we turn off general relativistic effects where they are included
(in the ray tracing and M1 codes).

We place a stationary observer at $r=500$~km.
Taking advantage of the spherical symmetry, we sample the distribution function
$f(\tilde{\varepsilon},\cos A,B)$ with ray tracing
over a uniform grid of 40 points in energy $\tilde{\varepsilon}\in(0,100)$
and 80 points in angle $\cos A\in(0.9,1)$, holding fixed $B=\pi$.
With these samples we compute energy luminosity from the
radial momentum density $\tilde{H}^r$,
using the midpoint rule to convert the integral in Eqn.~\ref{eqn:Ha} to a sum:
\begin{equation}
  \label{eqn:L_minkowski}
  L =
  C \frac{\Delta}{(2\pi)^3} \sum\limits_{m=0}^{N-1} \varepsilon_m^3
  \cos A_m f_m,
\end{equation}
with $C=4\pi r^2$,
$\Delta \equiv 2\pi\Delta\tilde{\varepsilon}\Delta(\cos A)$,
$N \equiv N_A N_{\tilde{\varepsilon}}$,
$N_A$ and $N_{\tilde{\varepsilon}}$ the number of samples in angle and energy,
and $m$ labeling each ray.
We also compute average energy as a function of incoming angle
$\langle\tilde{\varepsilon}\rangle(\cos A)$ using
Eqns.~\ref{eqn:avg_eps_per_cosA}--\ref{eqn:J_per_cosA},
and the average energy of all neutrinos measured by our observer using
\begin{equation}
  \label{eqn:avg_eps_minkowski}
  \langle\tilde{\varepsilon}\rangle =
  \frac{1}{N_A}\sum\limits_{m=0}^{N_A-1}\langle\tilde{\varepsilon}\rangle(\cos A_m).
\end{equation}

\begin{figure}
  \resizebox{\columnwidth}{!}{\input{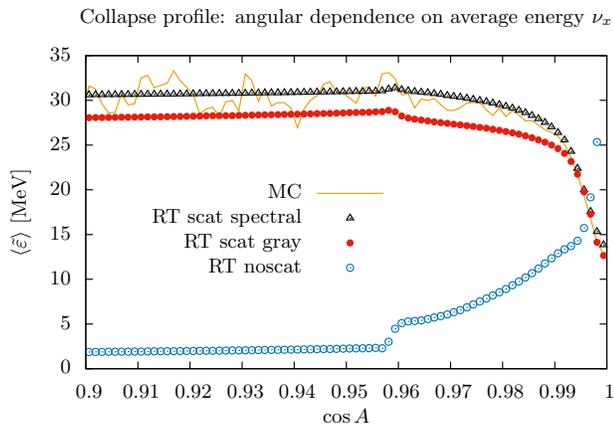}}
  \caption{Angular distribution of $\nu_x$ average energies
    $\langle\tilde{\varepsilon}\rangle(\cos A)$
    in the collapse profile test (described in Sec.~\ref{ssec:test_collapse}).
    The observer is at 500~km, and the shock at 150~km,
    so that the half-opening angle of the shock is $\cos A\approx0.954$.
    The four methods depicted are 1) a fiducial Monte Carlo calculation,
    ray tracing using the 2) spectral and 3) gray methods to estimate background
    fields for scattering, and 4) ray tracing neglecting scattering.
  }
  \label{fig:avg_eps_collapse}
\end{figure}

Fig.~\ref{fig:avg_eps_collapse} shows the distribution of average $\nu_x$
energies across incoming angles to this observer.
We show $\nu_x$ because they present the largest scattering effects:
they scatter through a thicker atmosphere outside their deep emission surface,
and their hotter spectrum experiences stronger modification due to the
$\varepsilon^2$ dependence of the scattering cross-section.
Against the fiducial Monte Carlo distribution, we show the
\emph{noscat} treatment, and the \emph{scat} treatment using both the
\emph{spectral} method and the \emph{gray} method.

As expected the average energies from both
the scattering envelope ($\cos A \lesssim 0.95$) and
the bright core ($\cos A\gtrsim 0.995 $)
are well characterized by the \emph{scat} treatments
and badly characterized by the \emph{noscat} treatment.
In this case for $\nu_x$ the major effect of elastic scattering
is to decrease the average energy of neutrinos coming from the core
and increase the average energy of neutrinos coming from the envelope.

Although not shown here, the angular distribution contributing to the
total number density is also strongly affected by elastic scattering.
Without elastic scattering, the central regions emitting 60\%
of the neutrinos for the species $\{\nu_e,\bar{\nu}_e,\nu_x\}$
have length scales $r_\nu\approx\{40,35,20\}$~km;
with elastic scattering the scales are
$r_\nu\approx\{50,45,35\}$~km.

The total luminosities and average energies
measured by the different treatments are
presented in Tab.~\ref{tab:collapse_rt_vs_mc}.
As expected, $\nu_e$ is least affected by scattering, and $\nu_x$ most.
In fact, without scattering, $\nu_x$ luminosities are overestimated by more than
two orders of magnitude, due to the steep increase in temperature with depth
in the inner core.
Though contributing only a fraction of the total luminosity,
the average energy of neutrinos scattered to the observer from the envelope
outside the shock, $\langle\varepsilon\rangle^{\rm sc}$,
is poorly characterized by the \emph{noscat} treatment for all species.
By contrast, both the \emph{gray} and \emph{spectral} \emph{scat} treatments
faithfully describe the high scattered energies from the envelope.

\begin{table}
  \caption{
    From the post-bounce collapse profile test (Sec.~\ref{ssec:test_collapse}),
    a comparison of total luminosities and average energies between
    the following methods:
    ray tracing ignoring elastic scattering `noscat',
    ray tracing with the gray scattering treatment `scat gray',
    ray tracing with the spectral scattering treatment `scat spectral', and
    a fiducial Monte Carlo transport evolution `MC'.
    We also show average energies of the scattering envelope
    $\langle\tilde{\varepsilon}\rangle^{\rm sc}$, which are estimated by eye
    from plots like Fig.~\ref{fig:avg_eps_collapse} at $\cos A < 0.95$.
    Luminosities have units $10^{52}\,{\rm erg}\,{\rm s}^{-1}$ and
    average energies have units MeV.
    The $\nu_x$ luminosities are per-species:
    multiply by four to get the total heavy-lepton neutrino luminosities.
  }
  \label{tab:collapse_rt_vs_mc}
  \begin{tabularx}{\columnwidth}{X r r r r}
    & {\bf noscat} & \,\,{\bf scat gray} & \,\,{\bf scat spectral} & \,\,{\bf MC} \\
    \hline
    $L_{\nu_e}$                                         &  3.53 & 3.14 & 3.04 & 3.48 \\
    $L_{\bar{\nu}_e}$                                   & 5.19 & 3.05 & 3.03 & 3.01 \\
    $L_{\nu_x}$                                         & 222.0  & 1.88 & 1.76 & 1.70 \\
    $\langle \tilde{\varepsilon}_{\nu_e} \rangle$               &  10.6 & 10.6 & 10.9 & 11.0 \\
    $\langle \tilde{\varepsilon}_{\bar{\nu}_e} \rangle$         & 14.2 & 13.0 & 13.8 & 13.7 \\
    $\langle \tilde{\varepsilon}_{\nu_x} \rangle$               & 47.8 & 16.0 & 17.3 & 16.2 \\
    $\langle \tilde{\varepsilon}_{\nu_e} \rangle^{\rm sc}$      & 4    & 14   & 17   & 16 \\
    $\langle \tilde{\varepsilon}_{\bar{\nu}_e} \rangle^{\rm sc}$& 4    & 24   & 20   & 18 \\
    $\langle \tilde{\varepsilon}_{\nu_x} \rangle^{\rm sc}$      & 2    & 28   & 31   & 27 \\
    \hline
  \end{tabularx}
\end{table}

We can make some quantitative sense of the scattered energies in
Fig.~\ref{fig:avg_eps_collapse} and Tab.~\ref{tab:collapse_rt_vs_mc}
using the solutions explored in the simple configurations above.
In particular, the average energy in the scattering envelope
($\cos A\lesssim 0.95$) is related to the spectrum in the direction
of the interior ($\cos A \sim 1$).
The relation may be derived by simplifying our realistic model
to that of a homogeneous matter profile and background scattering field.

By expanding the exponential of
Eqns.~\ref{eqn:fse_homogeneous_general} and \ref{eqn:fae_homogeneous_general}
as we did for Eqn.~\ref{eqn:fae_homogeneous_limits},
and furthermore factoring out the dominant energy dependence from the
cross-sections (i.e.\ $\chi\equiv\varepsilon^2 \zeta$
with $\zeta$ approximately constant), we have
\begin{align}
  \label{eqn:fae_homogeneous_thin_limit}
  f_{\rm AE}(\varepsilon)
  &= \varepsilon^2 \zeta_a \,f^{\rm eq}(\varepsilon)\,s, \\
  \label{eqn:fse_homogeneous_thin_limit}
  f_{\rm SE}(\varepsilon)
  &= \varepsilon^2 \zeta_s \,\Phi(\varepsilon)\,s,
\end{align}
where $s$ is the length of scattering envelope passed through by the ray.
In the envelope the local temperature is low so $f^{\rm eq}/\Phi\ll 1$,
and also $\zeta_a/\zeta_s\ll 1$,
so $f_{\rm SE}$ strongly dominates over $f_{\rm AE}$.
The average energy of the scattered field measured by our observer
is therefore
\begin{equation}
  \label{eqn:avg_eps_sc}
  \langle \tilde{\varepsilon} \rangle^{\rm sc}
  = \frac{\int d\varepsilon\,\varepsilon^3 f}{\int d\varepsilon\,\varepsilon^2 f}
  \approx \frac{\int d\varepsilon\,\varepsilon^5 \Phi}{\int d\varepsilon\,\varepsilon^4 \Phi}.
\end{equation}
Note that we have taken the liberty here of identifying the fluid-frame energy
$\varepsilon$ with the Eulerian energy $\tilde{\varepsilon}$,
since infall velocities in the envelope are $\sim0.1c$,
and as Fig.~\ref{fig:avg_eps_doppler} indicates,
the Doppler shift will therefore introduce an error into our analysis of
$\sim10\%$.

The spectrum of the background field $\Phi(\varepsilon)$ is well-approximated
as the \emph{scat} solution for $\cos A\sim 1$, since that is the dominant
source direction for neutrinos.
And because the scattering envelope is optically thin at all energies,
we can assume the neutrino spectrum is essentially unchanged in its passage
through the envelope.

In order to estimate $\langle \tilde{\varepsilon} \rangle^{\rm sc}$ analytically,
we must write the background field $\Phi(\varepsilon)$ analytically.
Direct Fermi-Dirac fits using a temperature and chemical potential
representative of a physical neutrinosurface fair poorly since neutrinos of
different energies decouple from the matter at different radii, over which
the thermodynamic state varies substantially.
Phenomenological Fermi-Dirac fits work well;
but so do pinched spectral fits which are much simpler
\cite{keil2003-pinched_spectra,miri2016-sn_neutrinos}:
\begin{equation}
  \label{eqn:pinched_spectra}
  \Phi^{\rm pi}(\varepsilon) \propto
  \varepsilon^{\alpha-2}
  \exp\left(-(\alpha+1)\frac{\varepsilon}{\langle\varepsilon\rangle}\right),
\end{equation}
(where our definition differs from that of \cite{keil2003-pinched_spectra}
by a factor of
$\varepsilon^2$, since we define our distribution function to be dimensionless
in natural units $\{\hbar,c\}=1$).

Pinching parameters
(calculated by eye from the \emph{spectral} \emph{scat} method)
for the species
$\{\nu_e,\bar{\nu}_e,\nu_x\}$ are $\alpha\approx\{3.6,5.1,2.3\}$.
Energy moments of pinched spectra like those in Eqn.~\ref{eqn:avg_eps_sc}
have a simple analytic form so that Eqn.~\ref{eqn:avg_eps_sc} becomes
\begin{align}
  \label{eqn:pinched_avg_eps_sc_1}
  \langle\tilde{\varepsilon}\rangle^{\rm sc}
  &\approx \frac{\alpha+3}{\alpha+1} \langle\varepsilon\rangle, \\
  \label{eqn:pinched_avg_eps_sc_3}
  &\approx \{16,18,28\}\,~{\rm MeV},
\end{align}
again for $\{\nu_e,\bar{\nu}_e,\nu_x\}$ respectively.
These analytic predictions agree with the average energies of the scattering
envelope to approximately 10\%
of all of the treatments including elastic scattering presented
in Tab.~\ref{tab:collapse_rt_vs_mc}, except for
$\bar{\nu}_e$ in the \emph{gray} \emph{scat} treatment,
which deviates from our prediction by 30\%.
This agreement is excellent despite the drastic
simplifications used in our model.

The $\langle\tilde{\varepsilon}_{\bar{\nu}_e}\rangle^{\rm sc}$ prediction
in the \emph{gray} \emph{scat} treatment is approximately 30\%
larger than the the fiducial Monte Carlo estimate. This is
due to the large negative local chemical potential
$\eta_{\bar{\nu}_e}\sim-10$ in the scattering envelope.
As described in Sec.~\ref{ssec:sources_se},
we use $\eta_\nu$ in the \emph{gray} treatment to construct our
synthetic spectrum from total neutrino densities $J$ and $G$.
This error and our successful analysis using pinched spectra above,
points the way to future improvements to the \emph{gray} method:
making better assumptions about the spectra which are less sensitive
to local fluid quantities.

\section{Applications}
\label{sec:applications}

In this section we use the ray tracing code to
first calculate global measures
of the neutrino fields outside of the hypermassive neutron star and disk
formed in a binary neutron star merger simulated by
\cite{fouc2016-m1_nsns, fouc2016-m1_evolve_n},
and compare ray tracing results to those from the M1 transport simulation.
Second, we examine neutrino oscillations in this environment,
using results from ray tracing to include the effect of neutrino-neutrino
interactions on flavor evolution.

\subsection{Neutrinos from a Hypermassive Neutron Star Remnant}
\label{ssec:test_disk_comparison}
The merger of two neutron stars by gravitational wave emission produces a
postmerger configuration composed of a single neutron star surrounded by a disk.
Because of strong differential rotation and shock heating,
the remnant may temporarily avoid collapse to a black hole,
even if its mass exceeds the threshold of dynamical
instability for a rigidly rotating neutron star \cite{duez2009-review}.
These objects, called hypermassive neutron stars,
may avoid collapse for thousands of seconds
depending upon a number of physical factors including thermal pressure,
magnetic fields, and the microphysics of the fluid
\cite{rezz2015-two_winds,ravi2014-hmns_collapse}.

Such a configuration was modeled in \cite{fouc2016-m1_nsns} by
evolving fluid and spacetime through the
final inspiral and merger of two identical neutron stars of
isolated gravitational mass 1.2~$M_{\odot}$. The configuration
was simulated using a gray M1 transport scheme for the neutrinos, evolving the
energy density, number density, and energy flux in addition to the standard
fluid and metric variables to $\sim11\,{\rm ms}$ following merger
\cite{fouc2016-m1_evolve_n}.
The fluid was modeled using the LS220 equation of state \cite{latt1991-nuc_eos}.

\begin{figure}
  \includegraphics[width=\columnwidth]{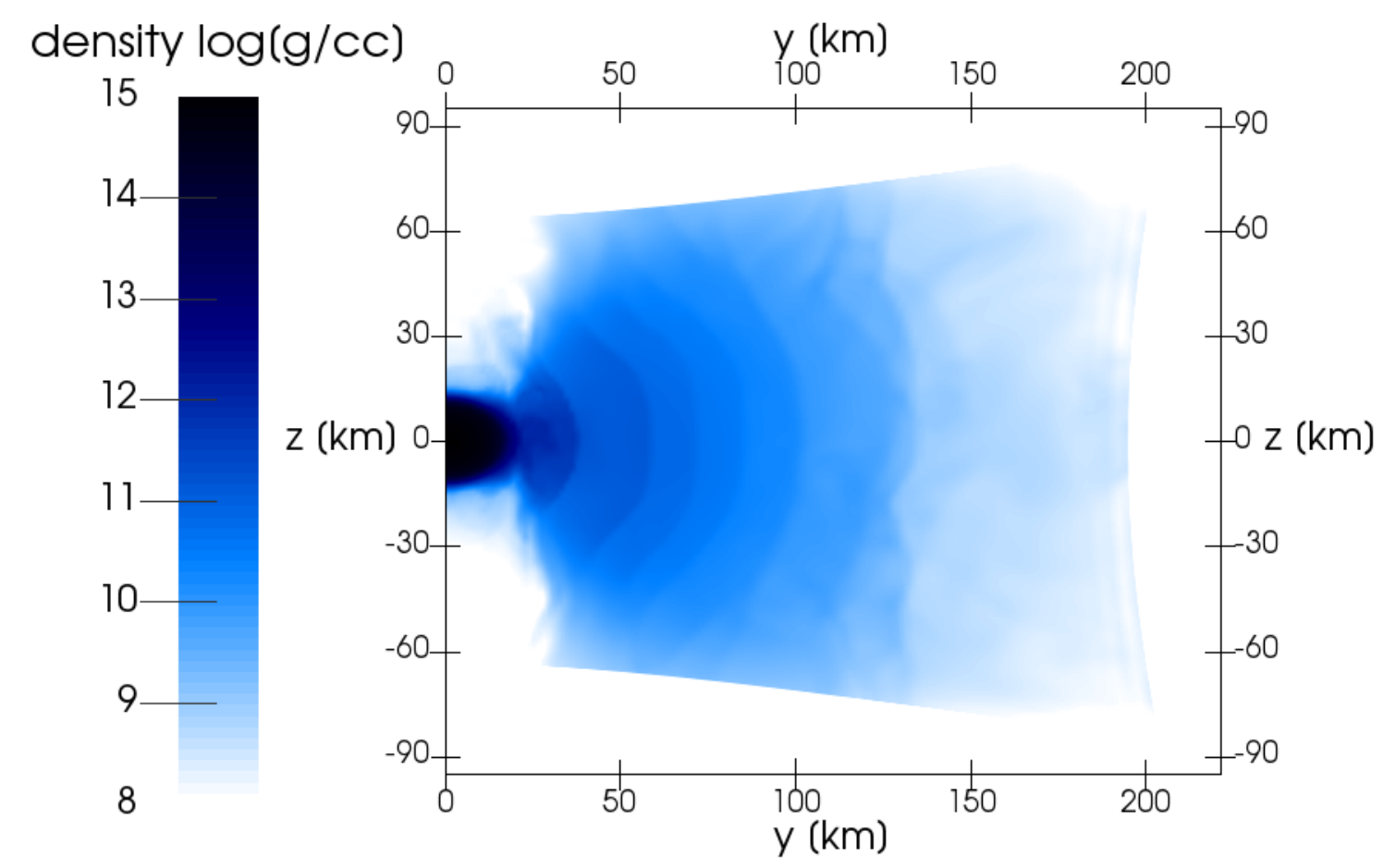}
  \caption{A meridional slice of density in the hypermassive neutron
    star and disk configuration (Sec.~\ref{ssec:test_disk_comparison}).
    The distorted rectangular boundaries are the boundaries of the grid
    used in the numerical simulation, which employs a coordinate mapping to
    concentrate points near the central object.}
  \label{fig:nsns_rho_merid}
\end{figure}

\begin{figure}
  \includegraphics[width=\columnwidth]{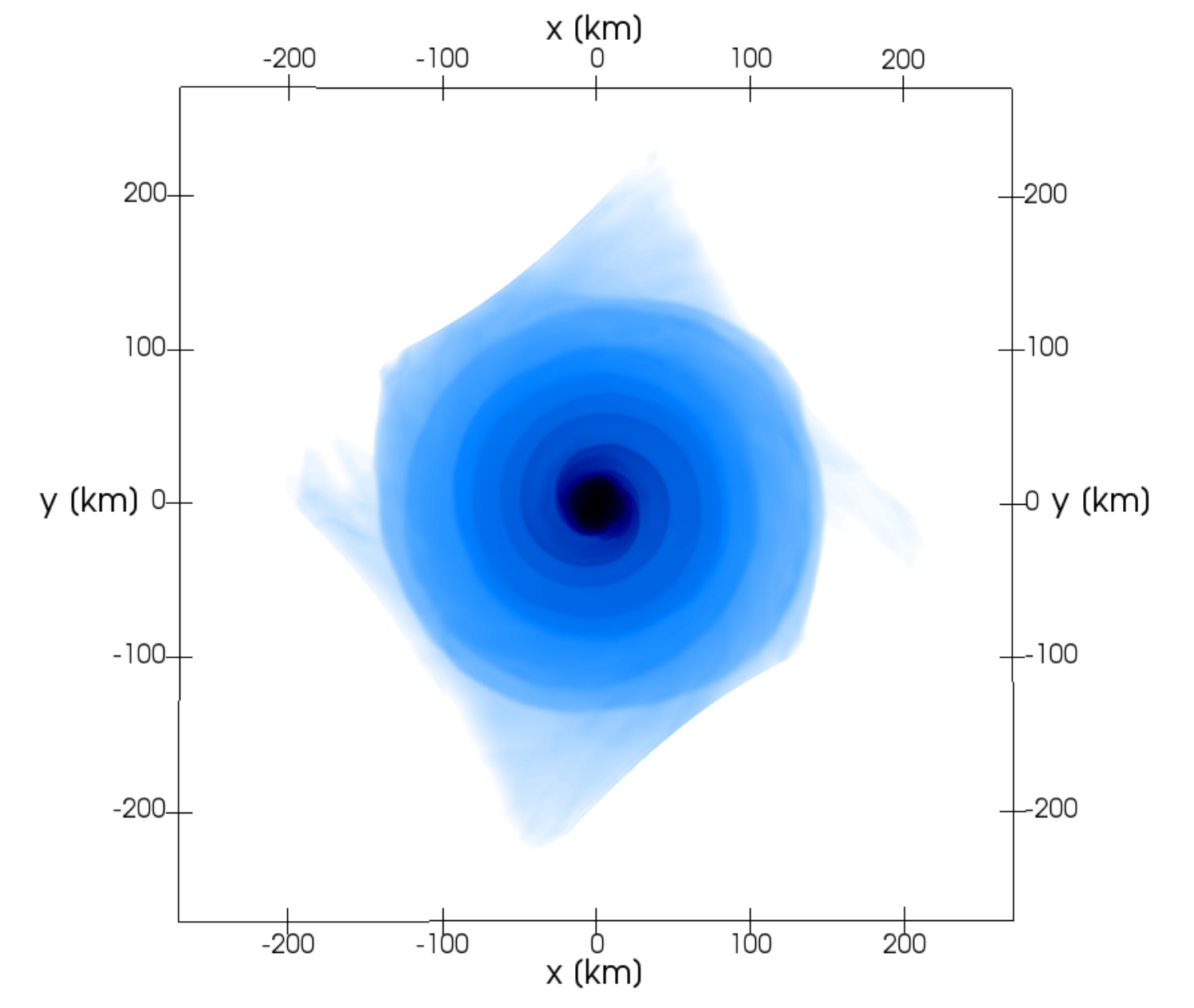}
  \caption{An equatorial slice of density in the hypermassive neutron
    star and disk configuration (Sec.~\ref{ssec:test_disk_comparison}).}
  \label{fig:nsns_rho_equat}
\end{figure}

\begin{figure}
  \includegraphics[width=\columnwidth]{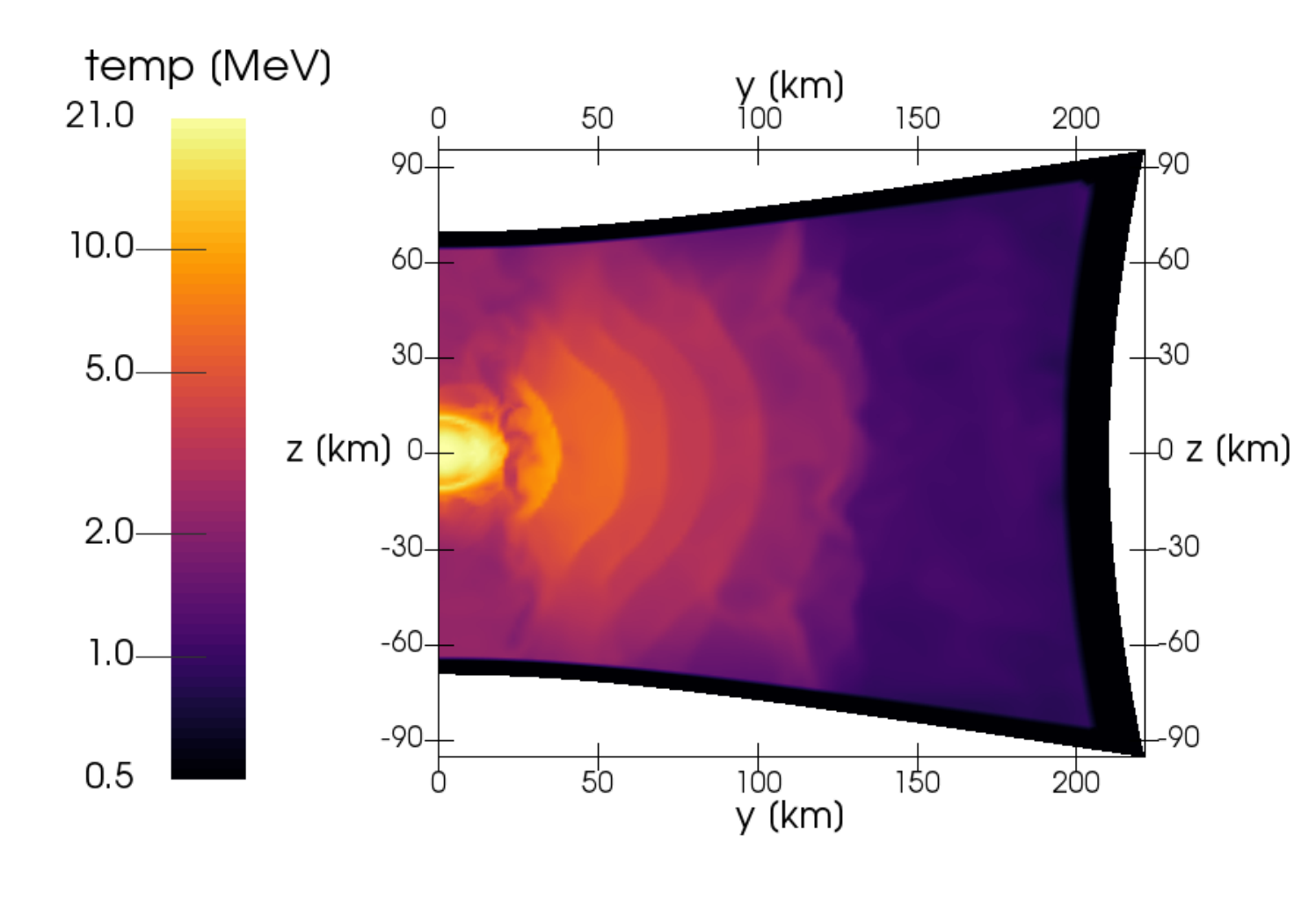}
  \caption{A meridional slice of temperature in the hypermassive neutron
    star and disk configuration (Sec.~\ref{ssec:test_disk_comparison}).}
  \label{fig:nsns_temp_merid}
\end{figure}

\begin{figure}
  \includegraphics[width=\columnwidth]{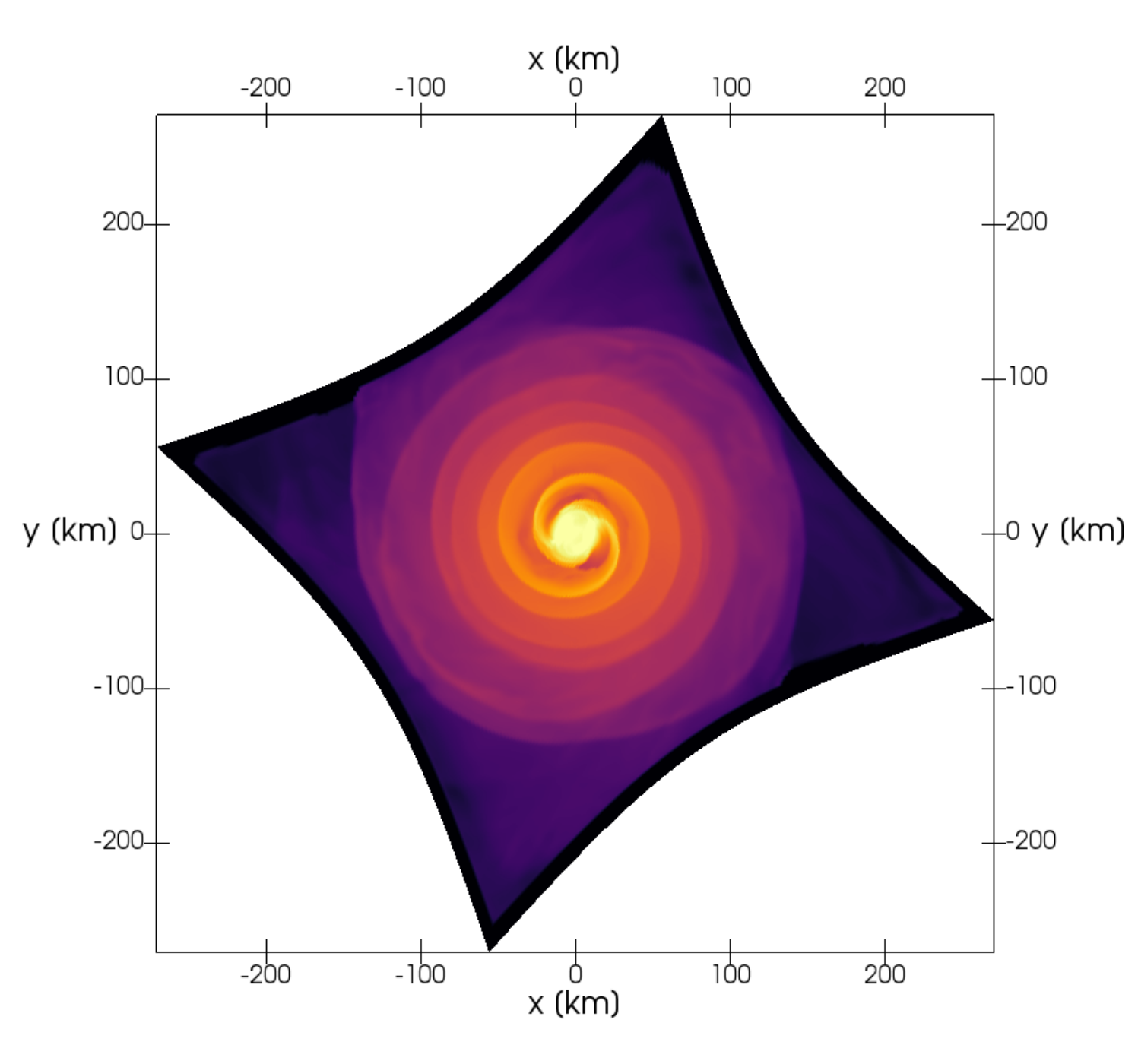}
  \caption{An equatorial slice of temperature in the hypermassive neutron
    star and disk configuration (Sec.~\ref{ssec:test_disk_comparison}).}
  \label{fig:nsns_temp_equat}
\end{figure}

We use a single time snapshot from that configuration
at $t=11\,{\rm ms}$ after merger.
In this way we approximate the system as stationary over a light-crossing time
of around 1~ms, which is far below the thermal timescale of the remnant.
Figs.~\ref{fig:nsns_rho_merid}--\ref{fig:nsns_temp_equat}
show slices of density and temperature from the finite difference
fluid and M1 radiation data.
These data were evolved on a rectilinear grid spanning approximately 400~km in
both the $x$ and $y$ directions, and 150~km in the $z$ direction in our
evolution coordinates.

We extrapolate the fluid and M1 radiation data from
the domain shown in Figs.~\ref{fig:nsns_rho_merid}--\ref{fig:nsns_temp_equat}
to a larger ray tracing domain by setting all fluid and M1 radiation variables
to their floor values outside the smaller domain.
This simple extrapolation is adequate for ray tracing since
the neutrinos are almost entirely free-streaming outside the smaller domain.
Since the metric data were evolved on this larger domain no extrapolation is
needed for them.
The larger domain is represented as a pseudospectral grid
composed of a sphere with concentric shells extending to $r\approx1400$~km.
Radial grid spacings are $\Delta r \approx 0.15$~km in the star
and $\Delta r \approx 2.5$~km in the disk,
with 12 cells spanning polar angles $\theta\in[0,\pi]$
and 24 cells spanning azimuthal angles $\phi\in[0,2\pi)$.
Though a pseudospectral representation of non-smooth hydrodynamic data
introduces some Gibbs-like oscillations in the variables,
we choose to use this representation instead of the mesh-refined
finite difference grid of the evolution because the pseudospectral
representation uses much less memory:
95~MB in pseuspectral vs.\ 3~GB in finite difference representation.
Obviously, memory loads of this order are not insurmountable;
but they would require some modifications to our
volume data interpolation infrastructure.
For the purposes of this analysis the pseudospectral representation is adequate.

Opacities are computed from the LS220 equation of state using
\lstinline{NuLib} and are stored as a table covering
energy, density, temperature, and composition ranges
spanning $\varepsilon\in[1,280.5]$~MeV logarithmically,
$\varrho\in[10^6,3.2\times10^{15}]\,{\rm g}\,{\rm cm}^{-3}$ logarithmically,
$T\in[0.05,150]$~MeV logarithmically, and 
$Y_e\in[0.035,0.55]$ linearly,
with grid extents $\{18,82,65,51\}$ respectively.

We place stationary (i.e.\ Eulerian) observers at fixed coordinate radius
$r=250$~km in the $y$-$z$ plane along an arc at $N_\theta$ positions
distributed linearly in $\cos\theta$ over the northern hemisphere
using $\theta_i=\cos^{-1}\left(1-i/(N_\theta-1)\right)$
for $i=0,1,2,\ldots N_\theta-1$.
We choose $N_\theta=30$.

Though the ray tracing sampling of $f$ is done in full general relativity,
in our calculation of moments at the observer positions,
we make the simplifying assumption of Minkowski spacetime.
The errors in our moments introduced by this assumption
may be estimated to be of order
$GM/r=4\,{\rm km}/250\,{\rm km}$
(with $G$ the gravitational constant and $M$ the central mass),
or $\sim 1\%$.

Each observer samples the distribution function over a uniform grid
in energy, cosine of polar angle, and azimuthal angle with extents
$\{N_{\tilde{\varepsilon}},N_A, N_B\}$ spanning the ranges
$\tilde{\varepsilon}\in(0,100)$~MeV,
$\cos A\in\left((\cos A)_{\rm min}, 1\right)$, and
$B\in[0,2\pi)$.
The integrals for fluxes of number and energy
(Eqns.~\ref{eqn:Ha}, \ref{eqn:Ka}, and \ref{eqn:J_H_S_eps_integrated})
become sums over all rays.
Taking these fluxes in the radial direction,
and using the midpoint rule to convert the integral to a sum, we have
\begin{align}
  \label{eqn:Kr_minkowski}
  \tilde{K}^r(\theta) &= \frac{\Delta}{(2\pi)^3}
  \sum\limits_{m=0}^{N-1}
  \tilde{\varepsilon}_m^2 \cos A_m f_m(\theta),\\
  \label{eqn:Hr_minkowski}
  \tilde{H}^r(\theta) &= \frac{\Delta}{(2\pi)^3}
  \sum\limits_{m=0}^{N-1}
  \tilde{\varepsilon}_m^3 \cos A_m f_m(\theta),
\end{align}
with $f_m(\theta) \equiv f(\theta;\tilde{\varepsilon}_m,\cos A_m, B_m)$,
$\Delta \equiv \Delta\tilde{\varepsilon}\, \Delta(\cos A)\, \Delta B$,
$N \equiv N_{\tilde{\varepsilon}}\, N_A\, N_B$,
and $m$ the index labeling each ray.
Average energies in the coordinate frame are given by
$\langle\tilde{\varepsilon}\rangle(\theta)=\tilde{H}^r(\theta)/\tilde{K}^r(\theta)$.
We choose $N_{\tilde{\varepsilon}}=30$, $N_A=150$, and $N_B=20$.
To maintain high angular resolution, we only sample rays that pass within
approximately 120~km of the star's center by setting $(\cos A)_{\rm min}=0.88$.

We also combine measurements from all observers to estimate
total luminosities and averages over the sky.
Since we have chosen an arc of observers isolated to the northern hemisphere and
the $y$-$z$ plane, we may extend these data to the full sky by assuming that
emission is azimuthally symmetric and reflection symmetric across the
equatorial plane. Figs.~\ref{fig:nsns_rho_merid}--\ref{fig:nsns_temp_equat}
indicate the approximate validity of these assumptions at 11~ms after merger.
Total luminosities are then computed as integrals of the radial fluxes
over $\cos\theta$.
Using the trapezoid rule, the number and energy luminosities become sums:
\begin{align}
  \label{eqn:luminosity_R}
  R &= C
  \left(\frac{1}{2}(\tilde{K}^r_0+\tilde{K}^r_{N_\theta-1})
  +\sum\limits_{n=1}^{N_\theta-2}\tilde{K}^r_n\right),\\
  \label{eqn:luminosity_L}
  L &= C
  \left(\frac{1}{2}(\tilde{H}^r_0+\tilde{H}^r_{N_\theta-1})
  +\sum\limits_{n=1}^{N_\theta-2}\tilde{H}^r_n\right),
\end{align}
with $C \equiv 2\pi r^2\, \Delta(\cos \theta)$,
$\tilde{K}^r_n \equiv \tilde{K}^r(\theta_n)$,
$\tilde{H}^r_n \equiv \tilde{H}^r(\theta_n)$,
and $n$ labeling each observer's position.
Average energies over the whole sky are then
$\langle\tilde{\varepsilon}\rangle=L/R$.
Note, for simplicity we use the coordinate radius $r$ in this expression even
though the earlier merger evolution did not necessarily produce
areal coordinates. The effect of this choice is to artificially scale the
ray tracing luminosities by some factor we believe to be very close to 1.
In future work we can correct this error by computing the proper area over
coordinate spheres at the observers' locations.

Tab.~\ref{tab:nsns_rt_vs_m1} compares the all-sky
luminosities and average energies from ray tracing and from
the M1 transport simulation,
which serves for qualitative comparison.
Even if M1 and ray tracing methods both provide faithful measurements
of all-sky luminosities, we expect some disagreement since the two
treatments differ fundamentally.
In addition to differences in transport methodologies,
the M1 fluxes are integrated over the outer boundary of the finite-difference grid
at radii ranging from 75~km to 200~km,
whereas the ray tracing fluxes are integrated over a sphere at radius 250~km,
introducing a time lag between some of the fluxes used in the measurements
of order one millisecond.
Additionally, uncertainties for the M1 simulation may be estimated from
comparisons between M1 methods to be around
15\% for energy luminosities and
10\% for average energies \cite[Sec.~A.6]{fouc2016-m1_evolve_n}.

\begin{table}
  \caption{
    Comparison of total luminosities and average energies
    of the hypermassive neutron star configuration
    (presented in Sec.~\ref{ssec:test_disk_comparison}).
    The methods are ray tracing using the \emph{noscat} method `noscat',
    ray tracing using the \emph{gray} \emph{scat} method `scat',
    and the M1 transport simulation `M1'.
    The ray tracing totals were computed from sums over
    observers placed in the $y$-$z$ plane.
    M1 values are taken from \cite[Figs. 7, 9, 10]{fouc2016-m1_evolve_n}.
    Energy luminosities have units $10^{52}\,{\rm erg}\,{\rm s}^{-1}$,
    number luminosities $10^{57}\,{\rm s}^{-1}$, and
    average energies MeV.
    The $\nu_x$ luminosities are per-species:
    multiply by four to get the total heavy-lepton neutrino luminosities.
  }
  \label{tab:nsns_rt_vs_m1}
  \begin{tabularx}{\columnwidth}{X r r r}
    & {\bf noscat} & \,\,{\bf scat} & \,\,{\bf M1} \\
    \hline
    $L_{\nu_e}$                                 & 5.87 & 5.40 & 5 \\
    $L_{\bar{\nu}_e}$                           & 9.70 & 10.7 & 12 \\
    $L_{\nu_x}$                                 & 23.6 & 12.5 & 12 \\
    $R_{\nu_e}-R_{\bar{\nu}_e}$                 & -0.91 & -1.68 & -2 \\
    $\langle \varepsilon_{\nu_e} \rangle$       & 12.7 & 12.0 & 12 \\
    $\langle \varepsilon_{\bar{\nu}_e} \rangle$ & 16.0 & 14.8 & 15 \\
    $\langle \varepsilon_{\nu_x} \rangle$       & 34.2 & 23.3 & 26 \\
    \hline
  \end{tabularx}
\end{table}

Because the model was evolved with a gray M1 transport scheme,
the only \emph{scat} method we can use in our ray tracing is the
\emph{gray} method.
As in the post-bounce configuration in Sec.~\ref{ssec:test_collapse},
the \emph{scat} treatment is more faithful than the \emph{noscat} treatment,
significantly so for the heavy-lepton neutrinos.
For example, with scattering turned off the ray tracing and M1 measurements
of $L_{\nu_x}$ disagree by 200\%;
but with scattering turned on agreement is within 10\%.
As the ray tracing and M1 measurements both show,
$R_{\bar{\nu}_e}$ dominates over $R_{\nu_e}$ significantly,
because the disrupted neutron star material is releptonizing.

\begin{figure}
  \resizebox{\columnwidth}{!}{\input{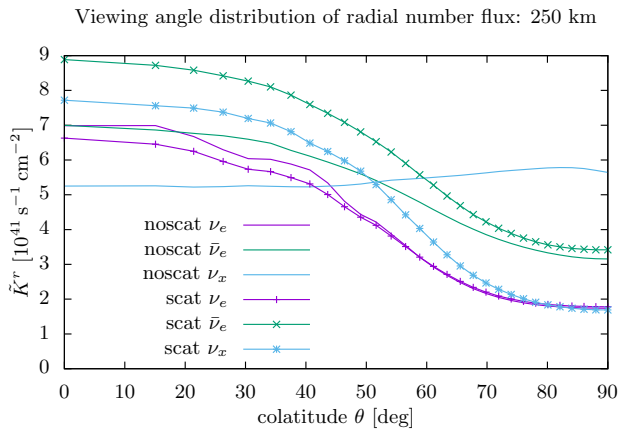}}
  \caption{Radial number fluxes, $\tilde{K}^r(\theta)$,
    as defined in Eqn.~\ref{eqn:Kr_minkowski},
    in the hypermassive neutron star configuration
    (presented in Sec.~\ref{ssec:test_disk_comparison}).
    Sampled for observers at fixed coordinate distances from the center
    of the star, with $r=250\,{\rm km}$.
  }
  \label{fig:nsns_theta_distrib_Kr}
\end{figure}

\begin{figure}
  \resizebox{\columnwidth}{!}{\input{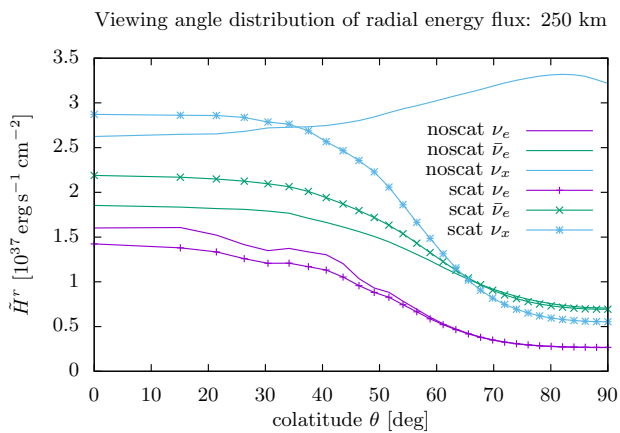}}
  \caption{Same as Fig.~\ref{fig:nsns_theta_distrib_Kr} but radial energy
    fluxes $\tilde{H}^r(\theta)$ as defined in Eqn.~\ref{eqn:Hr_minkowski}.
  }
  \label{fig:nsns_theta_distrib_Hr}
\end{figure}

The distributions of radial fluxes over observer position are shown
in Figs.~\ref{fig:nsns_theta_distrib_Kr} and \ref{fig:nsns_theta_distrib_Hr}.
With the \emph{noscat} treatment the $\nu_x$ luminosities are relatively
constant in $\theta$, because the disk is optically thin to $\nu_x$,
and most of the heavy-lepton neutrinos come from the hypermassive neutron star,
which is roughly spherical.
The slight upward trend in $\nu_x$ luminosities with $\theta$ may be due to
asymmetries in the fluid configuration,
inhomogeneous coordinate maps used in the hydrodynamics evolution,
or the fact that equatorial observers are closer to the disk's hot spiral arms
than are polar observers.
It is not due to the Doppler shift or relativistic beaming
from the rapid rotation of the star,
a hypothesis we tested by setting $u_i=0$ and $W=1$ in the ray
tracing equations.
Because of their larger absorption cross section,
the disk is not optically thin to $\nu_e$ and $\bar{\nu}_e$,
and observers at small angles within view of the hot hypermassive neutron star
measure the largest radial fluxes of these species.
When we turn on scattering, the disk is no longer transparent to $\nu_x$,
and $\nu_x$ luminosities present the same qualitative $\theta$-dependence
as that of the other species.

From Figs.~\ref{fig:nsns_theta_distrib_Kr} and \ref{fig:nsns_theta_distrib_Hr}
we also see that for observers near the poles ($\theta \sim 0^\circ$)
scattering generally increases fluxes of $\bar{\nu}_e$ and $\nu_x$
and decreases fluxes of $\nu_e$.
When scattering is turned on, all three fluxes experience a similar
loss of neutrinos from the central star.
But the $\bar{\nu}_e$ and $\nu_x$ fluxes experience a more dominant
gain of high energy neutrinos scattered by the disk back to the observer.
The $\nu_e$ fluxes, however, experience only a minor gain of neutrinos
from the disk, since $\nu_e$ presents a lower average energy,
and the scattering cross-section depends strongly on energy.
For observers in the equatorial plane ($\theta \sim 90^\circ$)
the effect of scattering is a decrease in energy fluxes for all species.
This is because more matter pollutes the equatorial regions than the
polar regions, causing the losses from the star to dominate
over the gains from the disk for all three species.

\begin{figure}
  \resizebox{\columnwidth}{!}{\input{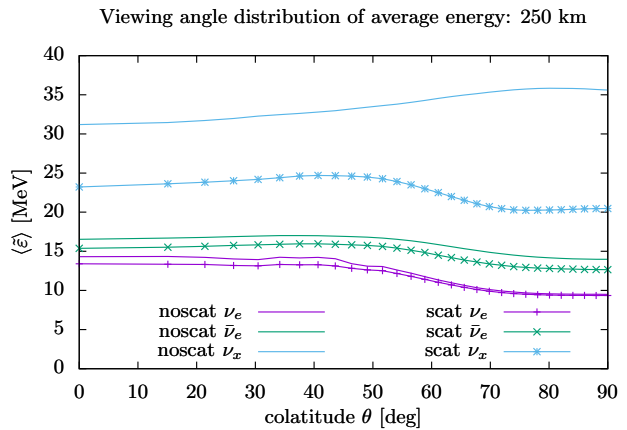}}
  \caption{Same as Fig.~\ref{fig:nsns_theta_distrib_Kr} but
    average energies of radial fluxes
    $\langle\tilde{\varepsilon}\rangle(\theta)=\tilde{H}^r(\theta)/\tilde{K}^r(\theta)$,
    as defined in Eqns.~\ref{eqn:Kr_minkowski} and \ref{eqn:Hr_minkowski}.
  }
  \label{fig:nsns_theta_distrib_avg_eps}
\end{figure}

Fig.~\ref{fig:nsns_theta_distrib_avg_eps} shows the distribution of
average energies of radial fluxes over observer positions.
With and without scattering, the average energies are highest
for observers near the polar axis, since polar observers get a
direct view of the hot hypermassive neutron star.
(This trend is unexpectedly reversed for $\nu_x$,
and may be due to asymetries in the disk, as discussed above.)
Scattering decreases average energies across all observer positions,
as it did in the post-bounce configuration, or any configuration of
a hot interior surrounded by a scattering envelope.
The strength of the effect of scattering on the average energies of
the different species is seen to follow the ranking
$\nu_x>\bar{\nu}_e>\nu_e$ due to two factors:
the average energies of the spectra obey the same ranking,
and the thicknesses of the different species' scattering envelopes
also obey the same ranking.
As we showed in the models with a homogeneous background scattering field,
the scattering contribution $f_{\rm SE}$ is
proportional to energy $\varepsilon^2$ and path length $s$
(Eqn.~\ref{eqn:fse_homogeneous_thin_limit}).

\begin{figure}
  \resizebox{\columnwidth}{!}{\input{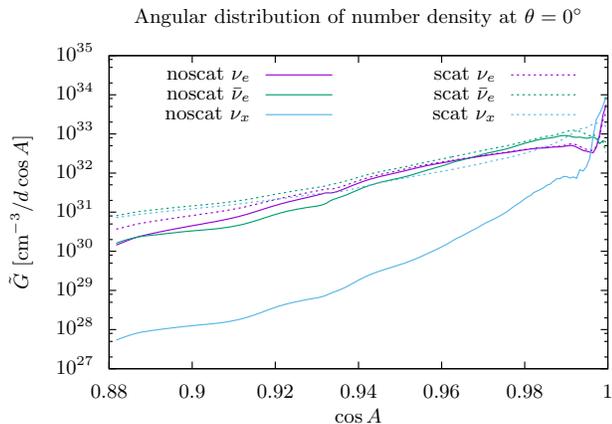}}
  \caption{Distribution of number density $\tilde{G}(\cos A)$ over incoming angle,
    defined in Eqn.~\ref{eqn:G_per_cosA_2}.
    The integral of this quantity over $d\cos A$ gives the total number density
    measured by this observer on the rotation axis.
    Volume data from the hypermassive neutron star configuration
    (presented in Sec.~\ref{ssec:test_disk_comparison}).}
  \label{fig:nsns_cosA_distrib_G_00deg}
\end{figure}

\begin{figure}
  \resizebox{\columnwidth}{!}{\input{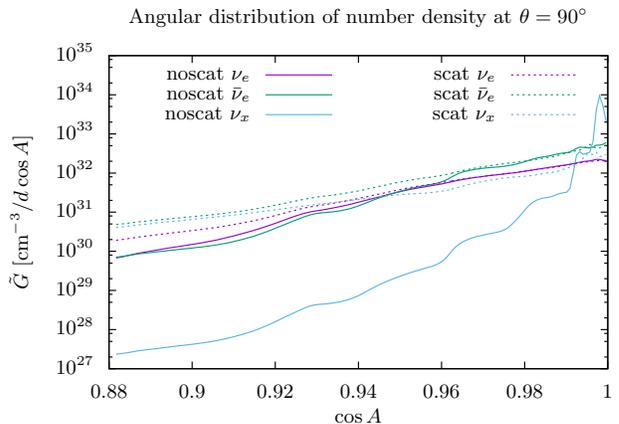}}
  \caption{Distribution of number density $\tilde{G}(\cos A)$ over incoming angle,
    as in Fig.~\ref{fig:nsns_cosA_distrib_G_00deg},
    but with the observer in the equatorial plane.}
  \label{fig:nsns_cosA_distrib_G_90deg}
\end{figure}

In Figs.~\ref{fig:nsns_cosA_distrib_G_00deg}
and \ref{fig:nsns_cosA_distrib_G_90deg}, we show the distributions of neutrino
number density over incoming polar angle $\cos A$
for the observer on the polar axis $\theta=0^\circ$,
and in the equatorial plane $\theta=90^\circ$.
This number density is defined
\begin{align}
  \label{eqn:G_per_cosA_2}
  \tilde{G}(\cos A) &=
  \frac{\Delta}{(2\pi)^3} \sum\limits_{m=0}^{N-1}\tilde{\varepsilon}_m^2 f_m(\cos A),
\end{align}
with $\Delta \equiv \Delta\tilde{\varepsilon} \Delta B$,
$N \equiv N_{\tilde{\varepsilon}} N_B$,
$f_m(\cos A) \equiv f(\tilde{\varepsilon}_m,\cos A,B_m)$,
and $m$ the index labeling each ray at polar angle $\cos A$.
The integral of this quantity over $d\cos A$ gives
the total number density $\tilde{G}$, according to Eqn.~\ref{eqn:G}.
As in the case of the collapse profile in Sec.~\ref{ssec:test_collapse},
the dominant effect of elastic scattering on $\tilde{G}(\cos A)$
is to spread the distribution out to larger angles, by generally decreasing
the number of neutrinos coming from the core while increasing the number
of neutrinos coming from the disk.

For the observer in the equatorial plane,
Fig.~\ref{fig:nsns_cosA_distrib_G_90deg}, the disk is so optically thick to
$\nu_e$ that there is very little difference between \emph{scat} and
\emph{noscat} treatments for incoming angles $\cos A \gtrsim 0.95$
corresponding to the volume inside of $r \lesssim 80$~km.
Also the stepped temperature gradient in the disk's spiral arms visible in
Figs.~\ref{fig:nsns_temp_merid} and \ref{fig:nsns_temp_equat} presents as a
stepped heavy-lepton neutrino number density distribution in
Fig.~\ref{fig:nsns_cosA_distrib_G_90deg} in the \emph{noscat} treatment,
since the energy emission due to $e^{-}e^{+}$ annihilation,
producing $\nu_x$, is especially sensitive to temperature, going as
$T^9$ \cite[Sec.~7]{burr2006-neutrino_opacities}.
This stepped distribution is not visible in $\nu_e$ and $\bar{\nu}_e$ emission,
since we ignore pair processes for these species due to the dominance of
absorption and emission processes, obeying a shallower temperature-dependence;
nor is it visible in the \emph{scat} treatment of $\nu_x$, since scattering
tends to smear the incoming angle of the emission;
nor is it visible for any species or treatment in
Fig.~\ref{fig:nsns_cosA_distrib_G_00deg}, since the integration over the
azimuthal angle $B$ averages out any spiral structure for the observer
on the polar axis.

\subsection{Neutrino Oscillations at High Neutrino Densities}
\label{ssec:V_nunu}
Here we examine the importance of elastic scattering in neutrino
flavor oscillation above the hypermassive neutron star--disk configuration
presented in Sec.~\ref{ssec:test_disk_comparison}.
Our treatment of the flavor evolution equation assumes flat space and small fluid
velocities.
In consequence we treat some of the gauge-dependent quantities inconsistently,
and we ignore potentially important features of relativistic flavor evolution
near compact objects \cite{yang2017-gr_nu_oscillations}.
We believe our treatment is sufficient, however, for the following qualitative
exploration.

The Boltzmann Equation (Eqn.~\ref{eqn:boltzmann}) is one limiting
form of the quantum kinetic equations for neutrinos
\cite{vlas2014-neutrino_qkes},
the limit where collisional mean free paths are much shorter
than oscillation lengths, i.e.\ $\varepsilon/\mathscr{K} \ll |H|^{-1}$ with
$H$ the Hamiltonian matrix describing coherent forward scattering interactions.
In this limit the neutrino density matrix takes the form
\begin{equation}
  \label{eqn:density_rho_0}
  \rho_0(\varepsilon,\ell_\mu) =
  \sum\limits_{\alpha}f^{\nu_\alpha}(\varepsilon,\ell_\mu)
  \ket{\nu_\alpha} \bra{\nu_\alpha},
\end{equation}
with $\ket{\nu_\alpha}$ the neutrino flavor eigenstates.
The neutrino state remains pure, or diagonal in the flavor basis.

But at the opposite limit, in the free-streaming regime,
coherence effects become important,
and the quantum kinetic equations take a Schr\"{o}dinger-like form:
\begin{equation}
  \label{eqn:schroedinger}
  i \frac{d}{ds} S = H S,
\end{equation}
where $ds$ is an interval of proper length traversed by the neutrino as
measured by our observer, and
$S$ is the neutrino flavor evolution matrix describing a mixed state
\begin{equation}
  \label{eq:density_evolution}
  \rho = S \, \rho_0 \, S^\dagger.
\end{equation}
In this limit, we may decompose the Hamiltonian matrix into vacuum,
matter, and neutrino contributions:
\begin{equation}
  \label{eqn:decompose_H}
  H = H_{\rm V} + H_e + H_{\nu\nu}.
\end{equation}
Explicit formulas for these matrices may be found in the oscillation literature,
e.g.\ \cite{duan2009-review},
but for clarity we only give their order-of-magnitude scales here:
\begin{align}
  \label{eqn:scale_H_V}
  |H_{\rm V}| &\sim \frac{\Delta m^2}{\varepsilon}, \\
  \label{eqn:scale_H_e}
  |H_e|       &\sim G_{\rm F}\left|n_{e^-}-n_{e^+}\right|, \\
  \label{eqn:scale_H_nunu}
  |H_{\nu\nu}|&\sim G_{\rm F} \left|G_{\nu_e}-G_{\bar{\nu}_e}\right|,
\end{align}
where $\Delta m^2$ is the mass-squared differences between neutrino mass
eigenstates,
$G_{\rm F}$ is the Fermi coupling constant,
$n_{e^-}$ and $n_{e^+}$ are the electron and positron number densities,
and $G_{\nu_e}$ and $G_{\bar{\nu}_e}$ the electron neutrino and antineutrino
number densities defined in Eqn.~\ref{eqn:G}.
In this statement of scale for $H_{\nu\nu}$ we only include the isotropic
components of the neutrino fields; for our calculations below, however, we
include the full angular distributions found via ray tracing.

In regimes in which $|H_{\nu\nu}|\ll|H_{\rm V},H_e|$,
the flavor evolution is locally soluble in a ray-by-ray method,
and reveals a rich and physically important phenonenology including
vacuum, solar, atmospheric, and terrestrial oscillations,
as well as oscillations in supernova envelopes.
Where neutrino densities are relatively high, however,
as in neutron star mergers, the problem must be solved globally.
To date, no method has been devised to handle this problem in systems lacking
spherical symmetry.

However, we can solve a similar but tractable problem along a single ray.
We assume that all the neutrino rays intersecting an event along a given test ray
have undergone the same flavor evolution history as that of the test ray:
i.e.\ the evolution matrix $S$ is the same for all rays sharing that event.
This is the so-called single-angle approximation,
which is widely used in the supernova oscillation literature
and has been shown to be qualitatively faithful in those environments.
We note, however, that recent studies have discovered 1) spherically symmetric
configurations for which a single-angle calculation produces qualitatively
different flavor evolution behavior than a full multi-angle calculation
\cite{vlas2018-multiangle},
and 2) azimuthally-symmetric configurations for which the single-angle
approximation masks certain instabilities
in the flavor evolution \cite{wu2017-fast_neutrino_conversions}.

The formalism of the single-angle approximation requires knowledge of
the unoscillated neutrino contribution to the Hamiltonian matrix
along a given test trajectory.
This is a function of the unoscillated neutrino self-interaction potential,
which for the $\alpha$-th flavor is:
\begin{equation}
  \label{eqn:V_nu_alpha}
  V_{\nu_\alpha,0}(\varepsilon,\ell_\mu) =
  \frac{\sqrt{2}G_{\rm F}}{(2\pi)^3} \varepsilon^2
  \oint d\Omega'(1-\omega')f^{\nu_\alpha}(\varepsilon,\ell'_\gamma),
\end{equation}
with the test ray propagating in direction $\ell_\mu$,
ambient rays propagating in directions $\ell'_\gamma$,
and the cosine of the angle between these given by
$\omega'=\psi^{\mu\gamma} \ell_\mu \ell'_\gamma$.
As in the moment equations (Eqns.~\ref{eqn:J}-\ref{eqn:Ka}),
the integral is taken over all directions $\ell'_\gamma$.

Implementing the single-angle approximation, we first use ray tracing to compute
the unoscillated neutrino self-interaction potentials $V_{\nu_\alpha,0}$
at several points along a test neutrino trajectory.
We then integrate the flavor evolution matrix $S$ along this test trajectory,
interpolating $V_{\nu_\alpha,0}$ to all points sampled by the integration.
And at each integration step we rescale $V_{\nu_\alpha,0}$ according to the
mixing specified by $S$.

If conditions are right, a resonant flavor transition
introducing significant mixing may occur very near
the point where neutrinos begin free-streaming.
The matter-neutrino resonance
\cite{malk2012-mnr_1,malk2015-mnr_2,malk2016-mnr_3},
can occur where the matter potential and
neutrino self-interaction potentials cancel,
or where $V_e + V_{\nu\nu,0}=0$, with
\begin{align}
  \label{eqn:V_e}
  V_e &= \sqrt{2} G_{\rm F} \frac{\varrho}{m_N} Y_e, \\
  \label{eqn:V_nunu0}
  V_{\nu\nu,0} &= \int d\varepsilon
  \big(V_{\nu_e,0}(\varepsilon)-V_{\bar{\nu}_e,0}(\varepsilon)\big),
\end{align}
with $\varrho$ the rest density and $m_{N}$ the nucleon mass.
The matter potential $V_e$ is always positive;
and far outside the accretion disks formed in neutron star mergers,
in which the disrupted neutron-star matter is rapidly releptonizing,
the total unoscillated neutrino self-interaction potential $V_{\nu\nu,0}$
is large and negative.

We examine this effect in the post-merger configuration already analyzed
in Sec.~\ref{ssec:test_disk_comparison}.
We calculate the self-interaction potential along a radial coordinate trajectory
making an angle $\theta=25^\circ$ with the rotation axis,
at $N_r=7$ positions $r\in\{30,50,82,135,223,368,608\}$~km.
We sample distribution functions at each of these positions over a grid
with extents $N_\varepsilon=30$, $N_A=200$, $N_B=30$;
energies range over $\varepsilon\in(0,100)$~MeV;
polar angles range over $\cos A \in ((\cos A)_{\rm min}, 1)$, with
$(\cos A)_{\rm min}\in\{-1,-1,-1,0.488,0.849,0.947,0.981\}$
for each of the $N_r$ positions;
and azimuthal angles range over the whole sky $B\in[0,2\pi)$.
We interpolate the logarithm of the self-interaction potentials
(Eqn.~\ref{eqn:V_nu_alpha})
in path length $\log r$ along the ray,
by fitting a 3rd order spline with continuous derivative,
across the $N_r$ observation points.
For $r>608$~km we extrapolate the self-interaction potentials using the
geometric fall-off of $r^{-4}$ applicable to the far-field limit of the
self-interaction potential \cite{malk2016-mnr_3}.

\begin{figure}
  \includegraphics[width=\columnwidth]{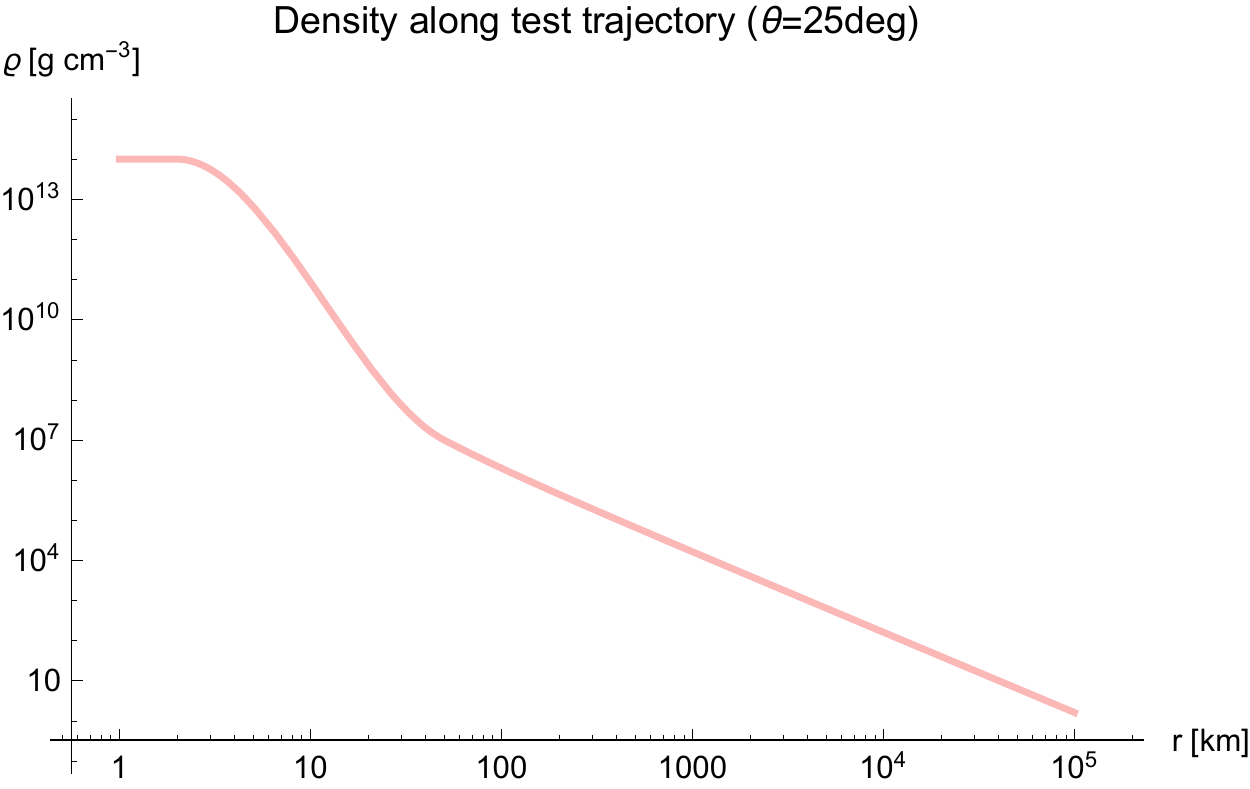}
  \caption{Assumed matter density along the neutrino test trajectory
    used to calculate the matter potential $V_e$.
    }
  \label{fig:test_traj_density}
\end{figure}

The matter potential along this test ray is from an analytic wind
model qualitatively consistent with the densities in the
simulated volume (i.e.\ inside $r\sim70$~km) and asymptoting to the $r^{-2}$
density field of a spherical steady-state wind with a constant asymptotic
velocity.
The density and velocity fields of a spherical steady-state wind
obey the continuity equation $\rho(r)v(r)r^2=\rho_1 v_1 r_1^2$,
with $\rho_1$ and $v_1$ the density and velocity measured at some fiducial
radius $r_1$.
For velocity field we choose a phenomenological wind model used in the
oscillation literature \cite{surm2005-nu_and_grb_outflows}
\begin{equation}
  v(r) = v_1 \left(1-\frac{R}{r_1}\right)^{-\beta}
  \left(1-\frac{R}{r}\right)^{\beta},
\end{equation}
with $R$ the wind launch radius and $\beta$ an acceleration parameter.
This yields the following density field
\begin{equation}
  \rho(r)=\rho_1 C\left[\frac{r_1}{R}\right]
  \left(1-\frac{R}{r}\right)^{-\beta}\left(\frac{R}{r}\right)^2,
\end{equation}
with $C[a]=a^{2-\beta}(a-1)^\beta$.
We use the parameters $R=10$~km, $r_1=50$~km, $\beta=2$,
and $\rho_1=10^7\,{\rm g}\,{\rm cm}^{-3}$.
Additionally we impose a density cap of
$\rho_{\rm max}=10^{14}\,{\rm g}\,{\rm cm}^{-3}$ inside the radius $r_0=5$~km,
and smoothly interpolate densities between $r_0$ and $r_1$ with a cubic polynomial,
enforcing $C^0$ and $C^1$ continuity at the transitions.
The rest density from this model is plotted in Fig.~\ref{fig:test_traj_density}.
To translate this density field to a matter potential, we assume a constant
electron fraction of $Y_e=0.5$, roughly consistent with the composition of the
matter in the disk's funnel \cite[Fig.~8]{fouc2016-m1_evolve_n}.

In fact, a spherical steady-state wind model, though providing an adequate
backdrop for the qualitative study presented in this section,
is less than ideal for this post-merger configuration,
most obviously because in the 10~ms since merger,
ejecta with the greatest velocities around $0.3c$
will have reached no further than $r_{\rm max} \sim 10^8$~cm.
Additionally, the true radial profile of the ejecta from this merger
(which our computational model does not follow)
will have many more features inside this radius, including shock jumps.
However, this model density profile is adequate as a backdrop to our study here
since we expect the remnant to present similar neutrino emission over a thermal
timescale of a few tens of milliseconds
while the matter field propagates out to larger radii.

\begin{figure}
  \includegraphics[width=\columnwidth]{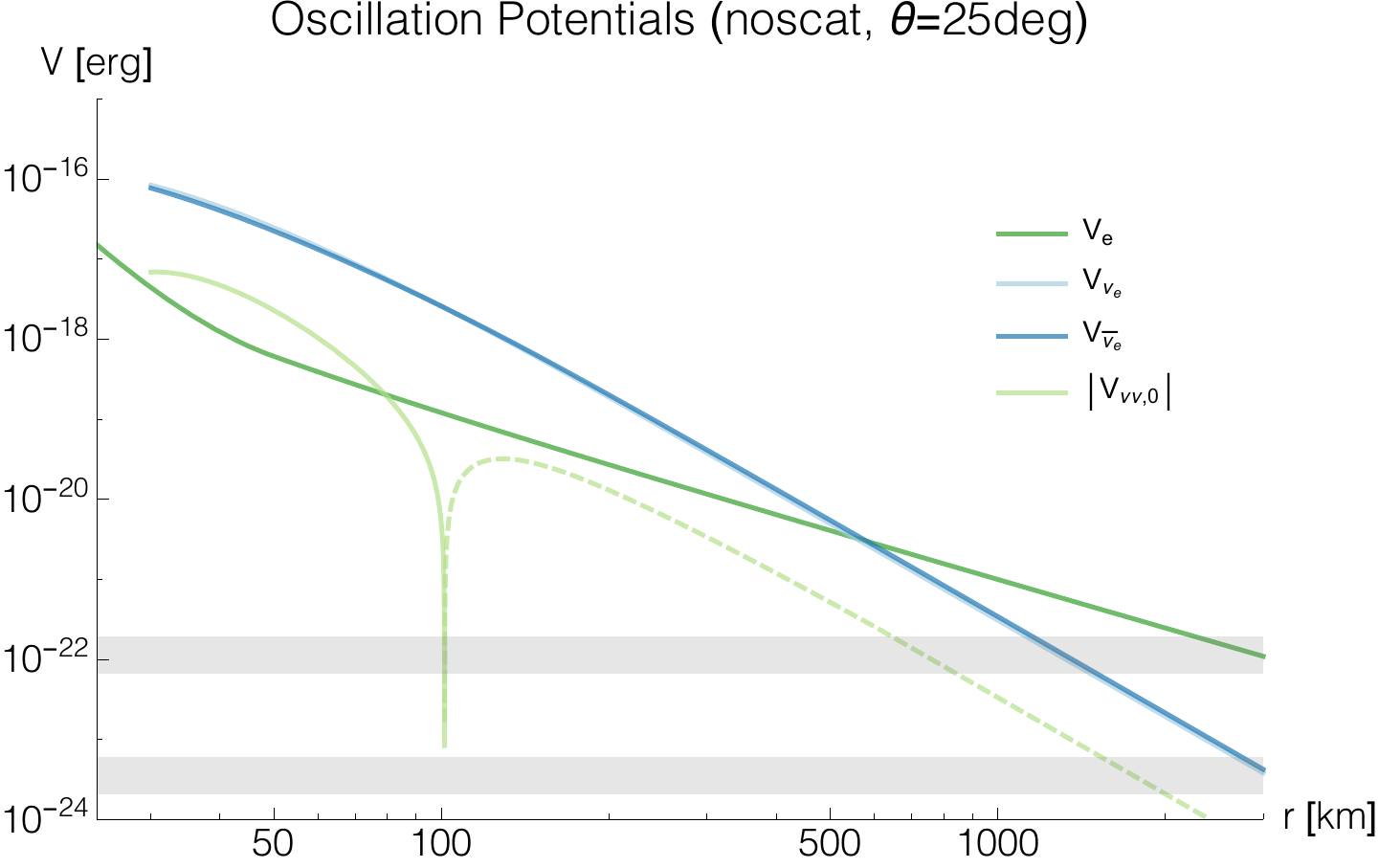}
  \caption{Neutrino oscillation potentials
    for the \emph{noscat} case
    along a test trajectory
    originating at the surface of the hypermassive neutron star, and
    proceeding outward along a radial coordinate ray with angle
    $\theta=25^{\circ}$ with respect to the polar axis.
    The trajectory is parameterized by the coordinate radius $r$.
    We plot the vacuum potential due to mass-squared differences (gray bands),
    the matter potential $V_e$ due to forward scattering on $e^-$ and $e^+$
    (dark green),
    and the self-interaction potential $V_{\nu\nu,0}$ due to forward
    scattering on ambient neutrinos (light green, solid where positive,
    dashed where negative);
    additionally we plot the $\nu_e$ and $\bar{\nu}_e$ components composing
    the self-interaction potential (light blue and dark blue respectively).
    The two gray bands at $V\sim10^{-22}$~erg and $V\sim10^{-24}$~erg
    indicating the two vacuum energy scales for 10--30~MeV neutrinos,
    set the positions of possible Mikheyev-Smirnov-Wolfenstein (MSW)
    or nutation resonances \cite{malk2012-mnr_1}.
    Inside the symmetric point at $r\sim100\,{\rm km}$ the total unoscillated
    self-interaction potential $V_{\nu\nu,0}$ is positive,
    i.e.\ $\nu_e$-dominated.
    }
  \label{fig:V_nunu-noscat}
\end{figure}

\begin{figure}
  \includegraphics[width=\columnwidth]{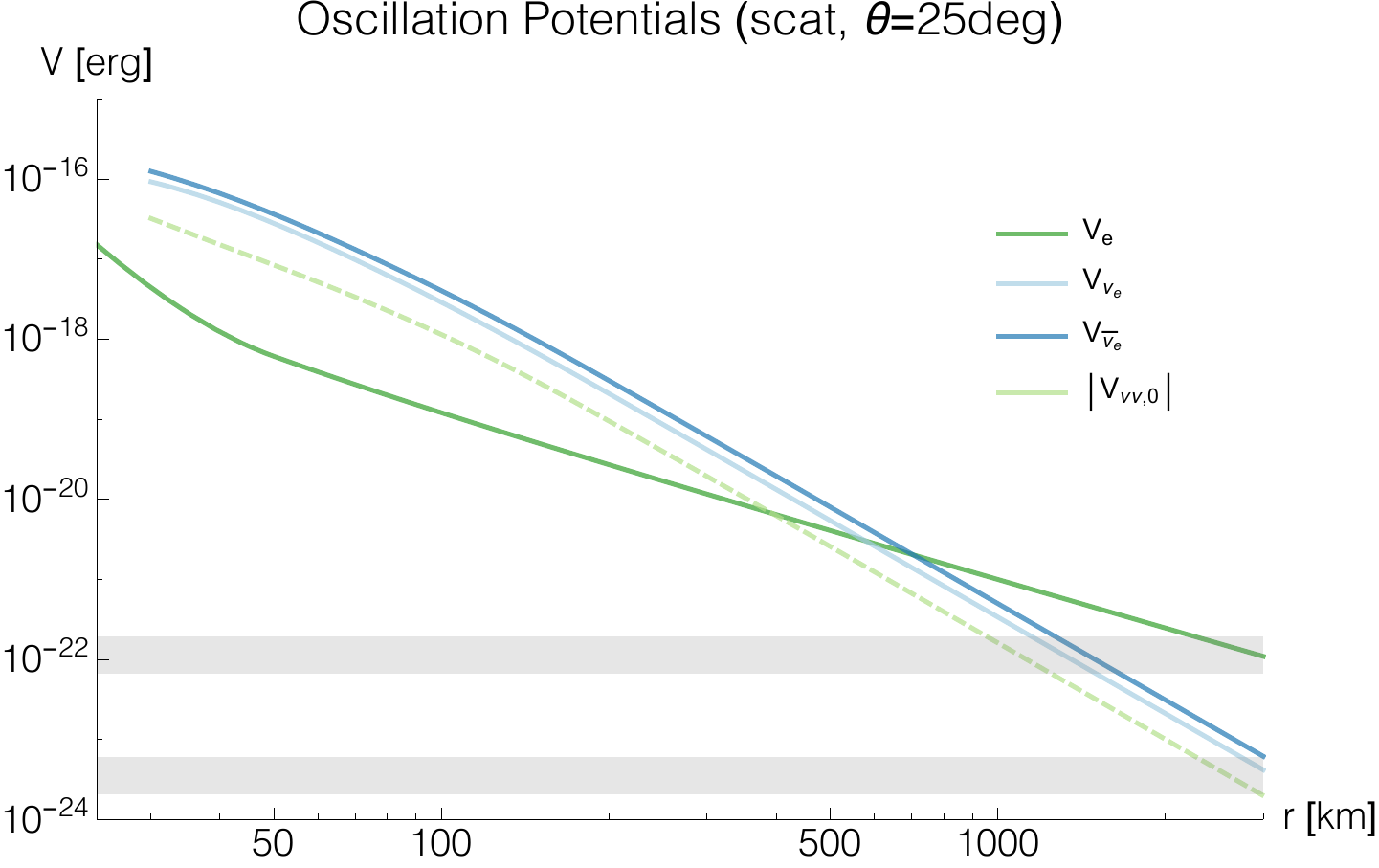}
  \caption{Same as Fig.~\ref{fig:V_nunu-noscat},
    but for the \emph{scat} case.
    Unlike the case with scattering turned off,
    here the total unoscillated self-interaction potential $V_{\nu\nu,0}$
    is everywhere negative, i.e.\ $\bar{\nu}_e$-dominated.
    }
  \label{fig:V_nunu-scat}
\end{figure}

In Figs.~\ref{fig:V_nunu-noscat} and \ref{fig:V_nunu-scat}
we show the total unoscillated self-interaction potential
and its contributions from $\nu_e$ and $\bar{\nu}_e$ for a test neutrino
moving out along the radial coordinate trajectory described above.
Fig.~\ref{fig:V_nunu-noscat} shows these terms for the \emph{noscat} treatment,
and Fig.~\ref{fig:V_nunu-scat} for the \emph{scat} treatment.
Obviously, including the effects of elastic scattering tends to increase the
$\bar{\nu}_e$ contribution relative to the $\nu_e$ contribution,
in this case causing the self-interaction potential to be negative
along the entire trajectory.
This effect may be predicted from Fig.~\ref{fig:nsns_cosA_distrib_G_00deg},
which is calculated for a qualitatively similar observer, in the vacated polar
funnel of the disk:
in the \emph{scat} treatment $G_{\bar{\nu}_e}$ dominates over $G_{\nu_e}$ at
all angles $\cos A \lesssim 0.998$;
whereas in the \emph{noscat} treatment $G_{\bar{\nu}_e}$ only dominates over
$G_{\nu_e}$ over a range of forward-peaked angles $\cos A \in (0.965,0.998)$.
An additional factor supporting this trend is that neutrinos from the most
forward-peaked angles $\cos A \sim 1$ have a suppressed effect on the
self-interaction potential due to the $(1-\cos A)$ term arising in the integral
of Eqn.~\ref{eqn:V_nu_alpha}.

\begin{figure}
  \resizebox{\columnwidth}{!}{\input{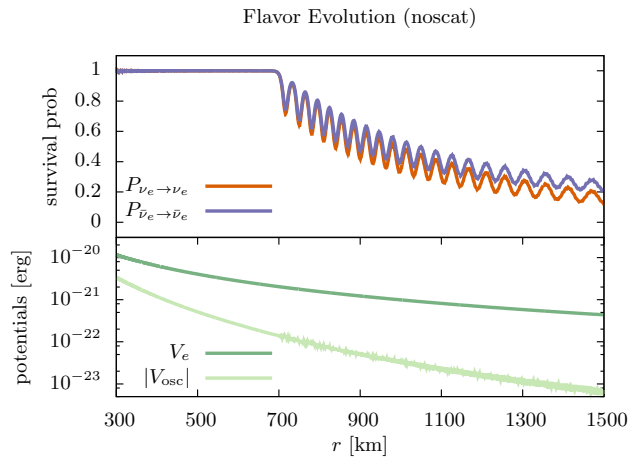}}
  \caption{Neutrino survival probabilities for the \emph{noscat} case.
    In the lower panel we show the matter potential $V_e$ from
    Fig.~\ref{fig:V_nunu-noscat},
    and the evolved self-interaction potential $V_{\rm osc}$ from
    Eqn.~\ref{eqn:V_osc}.
    In this case, the neutrinos undergo collective neutrino oscillation
    beginning around $r\sim700$~km,
    with both the electron neutrinos and antineutrinos converting to
    heavy-lepton neutrinos and antineutrinos almost completely.
  }
  \label{fig:survival-noscat}
\end{figure}

\begin{figure}
  \resizebox{\columnwidth}{!}{\input{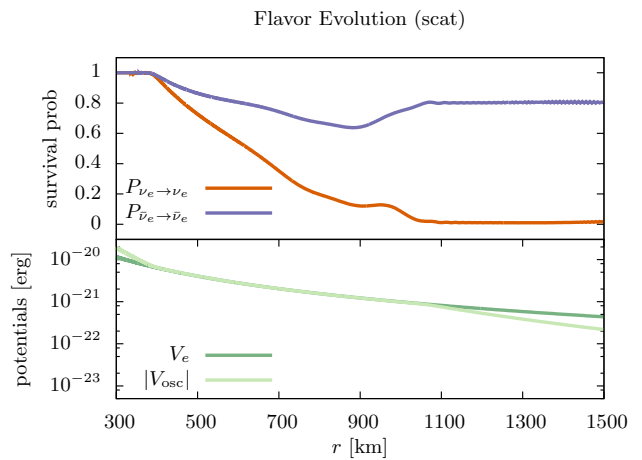}}
  \caption{Same as Fig.~\ref{fig:survival-noscat},
    but for the \emph{scat} case.
    In this case the neutrinos undergo a standard MNR transition
    beginning around $r\sim400$~km,
    with the electron neutrinos converting to heavy-lepton neutrinos almost
    completely,
    and the electron antineutrinos oscillating back to their original flavor
    after partially converting to heavy-lepton antineutrinos.
  }
  \label{fig:survival-scat}
\end{figure}

We also solve for the flavor evolution of this system,
integrating $S$ according to Eqn.~\ref{eqn:schroedinger},
as described in \cite{zhu2016-mnr_nsns_remnant}.
We show the survival probabilities for $\nu_e$ and $\bar{\nu}_e$
in Figs.~\ref{fig:survival-noscat} and \ref{fig:survival-scat},
comparing the \emph{noscat} and \emph{scat} treatments.
In these figures we also show the evolved self-interaction potential
$V_{\rm osc}$,
\begin{equation}
  \label{eqn:V_osc}
  V_{\rm osc} \equiv (H_{\nu\nu})_{ee}-{\rm Tr}(H_{\nu\nu})/3,
\end{equation}
to show how the neutrino interactions driving the oscillation evolve with flavor.
Note that $V_{\rm osc}$ is identical to $V_{\nu\nu,0}$ if no flavor
evolution takes place.
The survival probability is the probability that a neutrino, if measured,
will be found to be in its original flavor state.
The survival probabilities are computed at each point along the trajectory
from the absolute square of the diagonal terms of the flavor evolution matrix,
$P_{\nu_\alpha \to \nu_\alpha} = |S_{\alpha \alpha}|^2$.
When $P_{\nu_e \to \nu_e}$ decreases,
as can be seen for example in Fig.~\ref{fig:survival-scat} for $r>400$~km,
some of the $e$ neutrinos have oscillated into $\mu$ or $\tau$ neutrinos.

Fig.~\ref{fig:survival-noscat} shows the survival probabilities
from the calculation with elastic scattering turned off.
Using neutrino mixing angle $\theta_{12}$ and the inverted hierarchy
(the normal hierarchy gives qualitatively similar results)
electron neutrinos and antineutrinos start to oscillate around
700~km in the form of a collective neutrino oscillation,
causing both $\nu_e$ and $\bar{\nu}_e$ to convert to heavy lepton neutrinos
and antineutrinos respectively.
A similar effect was seen in \cite{fren2017-flavor_bns,tian2017-flavor_bns}.
Fig.~\ref{fig:survival-scat} shows the survival probabilities
of an otherwise identical calculation, but with elastic scattering turned on.
In this case, electron neutrinos and antineutrinos start to oscillate around
400~km in the form of a standard matter neutrino resonance:
at first both $\nu_e$ and $\bar{\nu}_e$ convert to heavy lepton neutrinos,
but as the transformation progresses, the $\bar{\nu}_e$ partially
return to their original flavor.

Note that during the latter part of the matter neutrino resonance
depicted in Fig.~\ref{fig:survival-scat},
where the self-interaction potential approaches the vacuum scale,
the survival probabilities show some small-scale oscillations
different than the standard matter neutrino resonance
introduced in \cite{malk2015-mnr_2}.
This occurs because the self-interaction potential and matter potential
fall close to the vacuum potential scale.

The collective neutrino oscillation occuring in the \emph{noscat} case
(Fig.~\ref{fig:survival-noscat})
not only produces a very different flavor mixture,
the transformation also starts further from the remnant,
and it extends much further before it completes.
For these potentials along this test trajectory,
we see that a matter-neutrino resonance only occurs in the \emph{scat} case;
though with a slightly higher matter potential it could occur in both cases,
closer to the disk in the \emph{noscat} case than the \emph{scat} case.
For this particular test trajectory we only see a standard and not a symmetric
matter neutrino resonance \cite{malk2016-mnr_3,vaan2016-uncovering_mnr}.
As we can see from these calculations, the final outcome of the flavor
transformation is quite different in the two scenarios.

\section{Conclusions}
\label{sec:conclusions}
We have introduced a new general relativistic ray tracing method to compute
neutrino distribution functions around compact objects in dynamical configurations,
and which incorporates the effects of elastic scattering for the first time
within a ray tracing framework.
Elastic scattering of neutrinos into and out of each ray is included in our
method by using estimates of the background neutrino fields from an
M1 transport simulation.
To capture the energy spectrum of the background field, we have described a
\emph{spectral} method which uses neutrino energy densities over multiple
energy groups as input, and a
\emph{gray} method which uses neutrino energy and number densities averaged
over all energies as input.
We have also successfully tested the ray tracing code with a comprehensive
battery of tests.

In our tests (Sec.~\ref{sec:tests}) we have confirmed that elastic scattering
plays a significant role in redistributing neutrino energy- and
angle-distributions in common compact-object configurations.
The largest effects are seen in $\nu_x$ distributions
and to a lesser extent $\bar{\nu}_e$,
with the dominant effect being a decrease in average energies from the central body,
and an increase in average energies from the scattering envelope.
More specifically, in the disk configuration formed by the merger of two neutron
stars (Sec.~\ref{ssec:test_disk_comparison}), elastic scattering causes
\begin{enumerate}
\item a decrease in average energies of neutrinos emerging from the remnant at
  all angles and for all species,
\item an increase in $\bar{\nu}_e$ and $\nu_x$ fluxes and a decrease in $\nu_e$
  fluxes viewed from along the rotation axis, and
\item a decrease in all species' fluxes viewed from the equatorial plane.
\end{enumerate}
Furthermore we find good agreement in overall number and energy luminosities
and average energies in comparisons with neutrino transport methods,
e.g.\ Monte Carlo in Sec.~\ref{ssec:test_collapse} and
M1 transport in Sec.~\ref{ssec:test_disk_comparison}.

We have also employed the ray tracing code to examine neutrino flavor
oscillations along one sample trajectory exiting the neutrino-dense environment
of the neutron star post-merger configuration (Sec.~\ref{ssec:V_nunu}).
The trajectory starts from 300~km, and moves out radially at
$25^\circ$ from the polar axis.
Along that trajectory, elastic scattering has the effect of increasing
the ratio of $\bar{\nu}_e$ relative to $\nu_e$.
This creates a negative self-interaction potential
which introduces a complete standard matter neutrino resonance transition
(see Fig.~\ref{fig:survival-scat}).
At about 400~km from the merger core,
both electron neutrinos and electron antineutrinos begin to transform.
At about 1200~km
$e$ neutrinos have almost completely converted to $\mu$ or $\tau$ neutrinos,
while the $e$ antineutrinos have returned back to their original flavor.
In an otherwise identical calculation, ignoring elastic scattering
causes the flavor transformation to be very different (see Fig.~\ref{fig:survival-noscat}).

This example demonstrates the importance of the physics of elastic scattering
in the phenomenon of neutrino flavor oscillation.
However, we avoid drawing general conclusions from this particular example,
since a single astrophysical configuration can present dramatically different
oscillation resonances along test trajectories emerging at different angles
\cite{zhu2016-mnr_nsns_remnant},
and since the matter neutrino resonance is extremely sensitive to a host
of parameters.

Finally, we propose the following improvements to the ray tracing code to make
it a more useful and robust astrophysical simulation tool:
\begin{enumerate}
\item Use the finite-difference hydrodynamics simulation grid to represent
  background fluid and neutrino variables instead of interpolating all
  input variables to the pseudo-spectral spacetime simulation grid in order
  to begin ray tracing.
  We have found that though the interpolation of fluid and neutrino variables
  to a lower-resolution pseudo-spectral grid saves computational memory, it
  introduces more costly problems, the foremost being Gibbs-like oscillations
  at shocks and discontinuities present in fluid fields.
\item Replace explicit with implicit time stepping in the integrations along
  each ray. We have found that the stability of the time-integration demands
  extremely small step sizes of the adaptive time-stepping algorithm, especially
  for higher-energy rays. Large errors are possible if time-stepping thresholds
  are not fine-tuned to each new configuration.
\item Improve the spectral assumptions made in the \emph{gray} method.
  The test of both \emph{gray} and \emph{spectral} methods against
  the fiducial Monte Carlo calculation presented in Sec.~\ref{ssec:test_collapse}
  indicated strong agreement for overall average energies for all species. 
  However, the average energy of $\bar{\nu}_e$ emerging from the envelope
  (which contributed only a tiny fraction to the total luminosity)
  differed between the \emph{gray} treatment and the Monte Carlo by 30\%.
  Agreement could be improved with better spectral assumptions, for example
  employing pinched spectra.
\end{enumerate}

\subsubsection*{Acknowledgements}
This work was supported in part by the National Science Foundation under Grant
No.\ \lstinline{PHY-1430152} (JINA Center for the Evolution of the Elements)
[MBD, YLZ, GCM];
by NASA through the Hubble Fellowship under grant \lstinline{51344.001-A},
awarded by the Space Telescope Science Institute,
which is operated  by the Association of Universities for Research in Astronomy Inc.\
for NASA under contract \lstinline{NAS 5-26555} [EO];
by NASA through the Einstein Postdoctoral Fellowship grant \lstinline{PF4-150122}
awarded by the Chandra X-ray Center,
which is operated by the Smithsonian Astrophysical Observatory for NASA
under contract \lstinline{NAS8-03060},
and through grant \lstinline{80NSSC18K0565} [FF];
through National Science Foundation Grant \lstinline{PHY-1402916} [MDD];
and through the U.S.\ Department of Energy, Office of Science,
Office of Nuclear Physics,
under award number \lstinline{DE-FG02-02ER41216} [GCM].
We thank James P.\ Kneller for his original flavor evolution code base.
We thank the Spectral Einstein Code collaboration
\footnote{\url{https://www.black-holes.org/code/SpEC.html}} for a powerful and
robust numerical relativity code base, and in particular
Lawrence E.\ Kidder, Daniel A.\ Hemberger, Fran\c{c}ois H\'{e}bert,
and Fatemeh Hossein Nouri for comments and help throughout.
We also thank Sherwood Richers for providing his neutrino transport code
\lstinline{Sedonu} for comparisons,
and Luke Roberts for helpful discussions at the JINA-CEE Frontiers meeting 2017.

\appendix

\section{Definitions}
\label{sec:definitions}
We decompose the neutrino momentum into components parallel and orthogonal to
an observer's velocity $u_\beta$:
\begin{equation}
  \label{eqn:def_momentum_2}
  p_\beta = \varepsilon (u_\beta + \ell_\beta),
\end{equation}
with $u_\alpha\ell^\alpha=0$ and $\ell_\alpha\ell^\alpha=1$.

We use two possible fiducial observers to define the momentum
decomposition via Eqn.~\ref{eqn:def_momentum_2}: the Eulerian,
or normal observer $n_\mu=-\alpha \,\partial_\mu t$,
and the fluid, or comoving observer
$u^\mu=Wn^\mu+v_{\rm E}^\mu$.
Here $t$ is coordinate time and $\alpha$ is the lapse
in the standard 3+1 decomposition of the metric
\begin{equation}
  \label{eqn:adm_metric}
  \psi_{\mu\gamma} \rightarrow
  \left(
  \begin{matrix}
    -\alpha^2 + \beta^i \beta_i  & \beta_i \\
    \beta_j                      & g_{ij}
  \end{matrix}
  \right).
\end{equation}
We have also introduced
the fluid Lorentz factor $W=\alpha u^t$,
its Eulerian velocity $v_{\rm E}^\mu=g^\mu_\lambda u^\lambda$
(distinct from its coordinate velocity $v^\mu=u^\mu/u^t$),
and the projection tensor orthogonal to the normal observer
$g^\mu_\lambda=\psi^\mu_\lambda+n^\mu n_\lambda$.

We specify the neutrino direction in the observer's frame
with two spherical polar angles
($a$,$b$) with respect to the simulation cartesian coordinates
\begin{align}
  \ell_\alpha &\rightarrow q(s,\Omega_i), \\
  \Omega_i &\rightarrow (\sin a \cos b,\sin a\sin b,\cos a),
\end{align}
or alternatively the two spherical polar angles
($A$,$B$) with respect to rotated coordinates
\begin{equation}
  \label{eqn:def_direction_primed_2}
  \Omega_{i'} \rightarrow  (\sin A \cos B,\sin A\sin B,\cos A).
\end{equation}
The two coordinate systems are related by a standard Euler rotation of first
$\phi$ about the $z$-axis, then $\theta$ about the rotated $y$-axis,
with $\phi$ and $\theta$ the azimuthal and polar position of the observer.
Expressed algebraically:
\begin{align}
  \label{eqn:rotation_connection}
  \Omega_x &= \Omega_{x'}\cos\theta\cos\phi-\Omega_{y'}\sin\phi+\Omega_{z'}\sin\theta\cos\phi,\nonumber\\
  \Omega_y &= \Omega_{x'}\cos\theta\sin\phi+\Omega_{y'}\cos\phi+\Omega_{z'}\sin\theta\sin\phi,\nonumber\\
  \Omega_z &= \Omega_{x'}\sin\theta+\Omega_{z'}\cos\theta.
\end{align}

The scale factors $q$ and $s$ are functions of the neutrino
direction $\Omega_i$, the observer's velocity $u_\alpha$,
and the spacetime metric.
In the case of a fluid observer, specified by an arbitrary $W$ and $u_i$,
$q$ and $s$ are given by
\begin{align}
  \label{eqn:q_general}
  q
  &= W\alpha\Bigg(2\beta^i\Omega_i W^2(\beta^i\Omega_i-1) \nonumber\\
  &\qquad\qquad - 2\Omega_i u_j g^{ij} W \alpha(\beta^i\Omega_i-1)\nonumber\\
  &\qquad\qquad + \alpha^2\left((\Omega_i u_j g^{ij})^2+\Omega_i\Omega_j g^{ij}W^2\right)\Bigg)^{-1/2}\\
  \label{eqn:s_general}
  s
  &= \beta^i\Omega_i -\Omega_i u_j g^{ij}.
\end{align}
In the case of an Eulerian observer, $W=1$ and $u_i=0$,
and these expressions simplify considerably:
\begin{align}
  \label{eqn:q_eulerian}
  \tilde{q}
  &= \alpha\left(2\beta^i\Omega_i(\beta^i\Omega_i-1)+\alpha^2\Omega_i\Omega_j g^{ij}\right)^{-1/2}\\
  \label{eqn:s_eulerian}
  \tilde{s}
  &= \beta^i\Omega_i.
\end{align}
In the even simpler case of Minkowski spacetime, these expressions
reduce to $q=W/(1+s)$, $s=-\Omega_i u^i$,
$\tilde{q}=1$, $\tilde{s}=0$.

In addition to the fluid velocity and Lorentz factor, described above, the
other fluid state variables we use from our hydrodynamic simulations
are rest density $\varrho=m_b n_b$, temperature $T$, and electron fraction
\begin{equation}
  Y_e = \frac{n_{e^-} - n_{e^+}}{n_b},
\end{equation}
where $n_{e^-}$, $n_{e^+}$, and $n_b$ are the number densities of electrons,
positrons, and baryons, and $m_b$ is the average baryon mass.

\section{Source Terms}
\label{sec:source_terms}
Here we present the sources comprising the right hand side of the
Boltzmann Equation (Eqn.~\ref{eqn:boltzmann}). The sources for the neutrino
distribution function $f(x^\alpha;p_\beta)$ arise from collision processes
producing, removing, or scattering to/from that point in phase space.
The weak interaction rates for each process involve integrals of
the neutrino distribution function $f(x^\alpha;p'_\beta)$ and that of the
antineutrino $\bar{f}(x^\alpha;p'_\beta)$ over a momentum volume
$dP'$ (Eqn.~\ref{eqn:dP}).

We follow
\cite{brue1985-core_collapse} and \cite[Sec.~4]{shib2011-truncated_moment}
by separating these processes into four categories:
\begin{equation}
  \label{eqn:four_sources}
  C[f] \equiv C_{\rm AE} + C_{\rm SE} + C_{\rm SI} + C_{\rm PP},
\end{equation}
representing charged-current absorption and emission, elastic scattering,
inelastic scattering, and the thermal pair processes of annihilation and
production. In this work, however, we only treat absorption/emission
and elastic scattering.
We seek to write each collision source linear in $f$:
\begin{align}
  \label{eqn:sources_cae}
  C_{\rm AE}
  &= \mathscr{E}_{\rm AE} - \mathscr{K}_{\rm AE}\, f, \\
  \label{eqn:sources_cse}
  C_{\rm SE}
  &= \mathscr{E}_{\rm SE} - \mathscr{K}_{\rm SE}\, f.
\end{align}

Each term
is computed by summing the weak interaction rates of the processes from the
relevant category given in Tab.~\ref{tab:neutrino_processes}.
We compute these rates in the rest frame of the fluid,
but because they are spacetime invariants
they take the same numerical value in any frame of reference
and are completely independent of our choice of fiducial observer $u^\alpha$
(see discussion around Eqn.~\ref{eqn:boltzmann_linear}).

We compute our rates using the open source neutrino interaction
library \lstinline{NuLib} \cite{ocon2015-gr1d_with_nu}.
We compile a table of sources defined over the four dimensions of
density, temperature, electron-fraction, and neutrino energy, and
interpolate quad-linearly to the points sampled along each ray.

Each source term is unique to the neutrino or antineutrino species modeled,
and consists of a sum over all of the processes contributing to that
category of interaction. For example:
$C^{\nu_e}_{{\rm AE}} = \sum_i\,C^{\nu_e}_{{\rm AE},i}$,
where $i$ labels the absorption/emission processes involving $\nu_e$
in Tab.~\ref{tab:neutrino_processes}.
By contrast $C^{\nu_x}_{{\rm AE}}$ is formally equal to zero.
However, in practice, \lstinline{NuLib} implements the thermal pair processes
via an effective emission/absorption term in order to avoid the need
to couple energy groups and species (see Sec.~\ref{ssec:sources_pp}).
This has been shown to work well for core-collapse supernovae
\cite{ocon2015-gr1d_with_nu}.

\subsection{Absorption and emission via charged current}
\label{ssec:sources_ae}
At neutrino and thermal energies well below the masses of the muon or tauon
($m_{\mu}\sim100\,{\rm MeV}$)
only charged current processes involving $\nu_e$ and $\bar{\nu}_e$ are allowed.
For each of these processes $i$ in Tab.~\ref{tab:neutrino_processes},
we may write an emission and absorption
coefficient as a function of the interaction cross-section
(e.g. \cite[Eqn.~A5]{brue1985-core_collapse}):
\begin{align}
  \label{eqn:sources_ae_1}
  C_{{\rm AE},i}
  &= \varepsilon j_i (1-f) - \varepsilon \chi_{a,i} f \\
  \label{eqn:sources_ae_2}
  &= \varepsilon j_i - (\varepsilon j_i+\varepsilon \chi_{a,i}) f,
\end{align}
where $j$ is the emissivity and $\chi_a$ the absorption opacity.
Both $j$ and $\chi_a$ have dimension ${\rm length}^{-1}$ and represent the
number of neutrinos emitted or absorbed per length traveled.
In radiation transport formulations using specific intensities instead of
distribution functions, an emissivity $\eta$ having dimension
${\rm energy}\,{\rm length}^{-3}\,{\rm time}^{-1}\,{\rm energy}^{-1}\,{\rm steradian}^{-1}$
is more commonly used.
The two are related by $j=(2\pi)^3\eta/\varepsilon^3$.
Note that for brevity we have suppressed the energy-dependence of
the terms $j$, $\chi_a$, and the distribution functions.

In the special case of radiative equilibrium we know that the source
term vanishes: an equal number of neutrinos are emitted from
and absorbed by the matter for any length traversed.
We also know in this case that the neutrino distribution function must be
$f^{\rm eq}$, the equilibrium Fermi-Dirac distribution function of
Eqn.~\ref{eqn:feq}.
With these facts we can rearrange Eqn.~\ref{eqn:sources_ae_1} to give us
Kirchoff's Law:
\begin{align}
  \label{eqn:kirchoffs_law}
  j_i &= \frac{\chi_{a,i}}{1-f^{\rm eq}}f^{\rm eq}, \\
  \label{eqn:stimulated_opacity}
  &= \chi^*_{a,i}\,f^{\rm eq},
\end{align}
where in Eqn.~\ref{eqn:stimulated_opacity} we have introduced the opacity
corrected for stimulated absorption, $\chi^*_{a,i}$.

Using these expressions and computing the sum over stimulated opacities
$\chi^*_{a,i}=\sum_i \chi^*_{a,i}$,
the invariant emissivity and opacity for absorption/emission are
\begin{align}
  \label{eqn:ae_emissivity_summed_2}
  \mathscr{E}_{{\rm AE}} &= \varepsilon \, \chi^*_a \, f^{\rm eq}, \\
  \label{eqn:ae_opacity_summed_2}
  \mathscr{K}_{{\rm AE}} &= \varepsilon \, \chi^*_a.
\end{align}
These expressions are equivalent to those in
Eqns.~\ref{eqn:ae_emissivity_summed} and \ref{eqn:ae_opacity_summed}.

We use the above treatment for $\nu_e$ and $\bar{\nu}_e$ only;
the $\mu$ and $\tau$ neutrinos and antineutrinos do not participate in
charged current absorption/emission interactions at these temperatures
and energies.
However we do use an effective stimulated absorption opacity $\chi^*_a$
for the heavy-lepton neutrinos, computed by \lstinline{NuLib} as
described in \cite{ocon2015-gr1d_with_nu} which follows
\cite{brue1985-core_collapse,burr2006-neutrino_opacities}.
This is described in App.~\ref{ssec:sources_pp}.

\subsection{Elastic scattering}
\label{ssec:sources_se}
Neutrino scattering on particles of mass much greater than $\varepsilon$
(i.e. nucleons and nuclei) is essentially iso-energetic. Following 
\cite[Eqn.~A8]{brue1985-core_collapse} or
\cite[Eqn.~4.20]{shib2011-truncated_moment} the collision term for the
$i$-th process takes the form
\begin{equation}
  \label{eqn:sources_se_1}
  C_{{\rm SE},i}(\ell_\alpha)
  = \frac{\varepsilon^3}{(2\pi)^3}
  \oint d\Omega' R_{{\rm SE},i}(\omega')
  \Big(f(\ell'_\beta) - f(\ell_\alpha)\Big)
\end{equation}
where $R_{\rm SE,i}(\omega')$ is the scattering kernel for the $i$-th process
from direction $\ell'_\beta$ to direction $\ell_\alpha$
having dimension ${\rm energy}^{-1}$,
and the cosine of the scattering angle is
$\omega' \equiv \psi^{\alpha\beta}\ell_\alpha\ell'_\beta$, with
$\psi^{\alpha\beta}$ the inverse of the spacetime metric.
Note that for brevity in Eqn.~\ref{eqn:sources_se_1} we have suppressed the
energy dependence of all of the terms.

It is customary to approximate the scattering kernel to linear order in
$\omega$ (as in~\cite[Eqn.~4.21]{shib2011-truncated_moment}):
\begin{equation}
  \label{eqn:se_kernel_expansion}
  R_{{\rm SE},i}(\varepsilon,\omega) \approx
  R_{{\rm SE},i}^0(\varepsilon) + \omega \, R_{{\rm SE},i}^1(\varepsilon).
\end{equation}
Using this definition and the moments defined in Eqns.~\ref{eqn:J}
and \ref{eqn:Ha}, and writing $\omega'=\ell_\alpha\ell'^\alpha$,
we expand Eqn.~\ref{eqn:sources_se_1} into four terms.
The term containing $\oint d\Omega' \ell'^\alpha$ vanishes by construction,
and the remaining three terms may be written in the form
\begin{equation}
  \label{eqn:sources_se_2}
  C_{{\rm SE},i}
  = \varepsilon \chi^0_i \Phi_i - \varepsilon \chi^0_i f,
\end{equation}
where we have introduced the scattering opacity $\chi^0_i$
for each of the elastic scattering processes $i$ in
Tab.~\ref{tab:neutrino_processes} and the background scattering field $\Phi$:
\begin{align}
  \label{eqn:opacity_s}
  \chi^0_i(\varepsilon)
  &= \frac{4\pi\varepsilon^2}{(2\pi)^3}R_{{\rm SE},i}^0(\varepsilon), \\
  \label{eqn:background_phi}
  \Phi_i(\varepsilon,\ell_\alpha)
  &= \frac{(2\pi)^3}{\varepsilon^3}\frac{1}{4\pi}
  \left(J(\varepsilon)+
  \frac{\chi^1_i(\varepsilon)}{\chi^0_i(\varepsilon)}
  \ell_\alpha H^\alpha(\varepsilon)\right),
\end{align}
and with $\chi^1_i$ defined
\begin{equation}
  \label{eqn:opacity_s_tr}
  \chi^1_i(\varepsilon)
  = \frac{4\pi\varepsilon^2}{(2\pi)^3}R_{{\rm SE},i}^1(\varepsilon).
\end{equation}

Note that $\chi^1_i/\chi^0_i$ in Eqn.~\ref{eqn:background_phi}
is the degree of non-isotropy in the scattering.
This term is roughly $-0.1$ for scattering on free neutrons,
$-0.2$ on free protons, and $1$ on heavy nuclei;
and at disk temperatures the composition is almost entirely free nucleons.
Therefore in our treatment, for simplicity,
we only retain the isotropic contribution to the background field:
\begin{equation}
  \Phi(\varepsilon)
  = \frac{(2\pi)^3}{\varepsilon^3}\frac{1}{4\pi}
  J(\varepsilon).
\end{equation}

Computing sums over the opacities
$\chi_s\equiv\sum_i\chi^0_i$ and
$\chi^1_s\equiv\sum_i\chi^1_i$,
the invariant emissivity and opacity for elastic scattering are
\begin{align}
  \label{eqn:se_emissivity_summed}
  \mathscr{E}_{\rm SE}
  &= \varepsilon \, \chi_s \, \Phi, \\
  \label{eqn:se_opacity_summed}
  \mathscr{K}_{\rm SE}
  &= \varepsilon \, \chi_s.
\end{align}

We have made the energy-dependence of the background scattering field
explicit in Eqn.~\ref{eqn:background_phi}.
When we compute $\Phi(\varepsilon)$ using $J(\varepsilon)$ and
$H^\alpha(\varepsilon)$ from a multi-group
M1 transport evolution, we call this the \emph{spectral} method.

However moment evolutions with multiple energy groups are still rare.
Most simulations employ a gray moment scheme,
evolving only the energy-integrated moments, $J$ and $H^\alpha$
and sometimes the number density $G$.
If such is the case we resort to the \emph{gray} method,
by reconstructing the energy-dependent source terms
from gray moments, assuming a diluted Fermi-Dirac spectrum.
We use the following procedure:
\begin{enumerate}
\item
  Interpolate the fluid temperature, $T$,
  equilibrium neutrino chemical potential, $\eta_\nu$,
  and evolved neutrino energy and number densities in the fluid frame,
  $J$ and $G$, from the simulation grid.
\item
  Compute the average neutrino energy in the fluid frame
  \begin{equation}
    \langle \varepsilon \rangle \equiv J/G.
  \end{equation}
\item
  Compute the neutrino spectral temperature, assuming equilibrium with the
  fluid
  \begin{equation}
    \label{eqn:spectral_temp}
    T_\nu = \langle \varepsilon \rangle
    \frac{\mathscr{F}_2(\eta_\nu)}{\mathscr{F}_3(\eta_\nu)},
  \end{equation}
  where $\mathscr{F}_b$ is the Fermi integral
  \begin{equation}
    \label{eqn:fermi_integral}
    \mathscr{F}_b(\eta) = \int_0^\infty dx \, x^b (1+e^{x-\eta})^{-1}.
  \end{equation}
  We implement the Fermi integrals using the analytical approximants from
  \cite{taka1978-beta_rates}.
\item
  Assume the background neutrino fields have the same total density
  as the evolved moments
  \begin{align}
    \label{eqn:J_from_gray}
    J(\varepsilon) &=
    J \frac{\varepsilon^3}{T_\nu^4 \,\mathscr{F}_3(\eta_\nu)}
    \big(1+\exp(\varepsilon/T_\nu-\eta_\nu)\big)^{-1}, \\
    \label{eqn:H_from_gray}
    H^\mu(\varepsilon) &=
    H^\mu \frac{\varepsilon^3}{T_\nu^4\, \mathscr{F}_3(\eta_\nu)}
    \big(1+\exp(\varepsilon/T_\nu-\eta_\nu)\big)^{-1}.
  \end{align}
\end{enumerate}

Note that to avoid floating point errors for very large negative $\eta$,
we employ the asymptotic form of Eqn.~\ref{eqn:fermi_integral}:
$\lim_{\eta\ll-1}\mathscr{F}_b(\eta) = b!\,e^\eta$.
Thus, for $\eta<-10$, we use the limiting forms of
Eqns.~\ref{eqn:spectral_temp},
\ref{eqn:J_from_gray},
and \ref{eqn:H_from_gray}:
\begin{align}
    T_\nu &= 
    \frac{\langle \varepsilon \rangle}{3}, \\
    J(\varepsilon) &=
    J \frac{\varepsilon^3}{6 T_\nu^4} e^{-\varepsilon/T_\nu}, \\
    H^\mu(\varepsilon) &=
    H^\mu \frac{\varepsilon^3}{6 T_\nu^4} e^{-\varepsilon/T_\nu}.
\end{align}

When we use the moments from Eqns.~\ref{eqn:J_from_gray} and
\ref{eqn:H_from_gray} in Eqn.~\ref{eqn:background_phi},
we call this the \emph{gray} method.

\subsection{Inelastic scattering}
\label{ssec:sources_si}
Neutrino scattering off of electrons is inelastic, changing the
magnitude and direction of the neutrino's momentum \cite{brue1985-core_collapse}.
In a supernova environment we expect inelastic scattering off of electrons
and nucleons to shift the neutrino spectra to lower energies, most noticeably
for heavy-lepton neutrinos \cite{thom2003-ccsne_neutrinos}.
A similar formalism to the above could be used to derive source terms
for inelastic scattering.
We save that for future work, pointing out here that inelastic scattering
treated this way is very sensitive to the energy-dependence of
the background field. In this work we take $C_{\rm SI}=0$.

\subsection{Thermal pair annihilation and production}
\label{ssec:sources_pp}
We do not include thermal pair processes within the standard pair-process
formalism; in other words we take $C_{\rm PP}=0$.
But as mentioned in App.~\ref{ssec:sources_ae}, we do include
these processes in an effective emission/absorption opacity for $\mu$
and $\tau$ neutrinos and antineutrinos.
This has been shown to work well for core-collapse supernovae
\cite{ocon2015-gr1d_with_nu}. Within \lstinline{NuLib}, we compute
the energy-dependent emissivity of the pair processes in
Tab.~\ref{tab:neutrino_processes}, ignoring final state neutrino blocking.
We then apply Kirchoff's law, Eqn.~\ref{eqn:kirchoffs_law},
to convert this to an effective absorption opacity.
We only use this effective opacity for heavy-lepton neutrinos
because the charged-current emission/absorption processes for $\nu_e$
and $\bar{\nu}_e$ dominate in the environments we study: early merger
remnants and supernovae.

\bibliography{references}

\end{document}